\documentclass[fleqn,usenatbib]{mnras}

\usepackage[T1]{fontenc}

\DeclareRobustCommand{\VAN}[3]{#2}
\let\VANthebibliography\thebibliography
\def\thebibliography{\DeclareRobustCommand{\VAN}[3]{##3}\VANthebibliography}

\usepackage{mathtools}
\usepackage{txfonts}
\usepackage{xcolor}
\usepackage{graphicx}
\usepackage{stfloats} 

\def\fado{{\sc FADO}}

\def\D4000{$D_{4000}$}

\newfont{\nlx}{cmssdc10 scaled 900}
\newcommand{\brem}[1]{\textcolor{black}{\nlx #1}}
\newfont{\hvss}{cmssdc10 scaled 1540}

\def\?{{\bf\color{red}?}}
\newcommand{\sextractor}{{\sc SExtractor}}
\newcommand{\cigale}{{\sc CIGALE}}
\newcommand{\pros}{{\sc Prospector}}

\title[Multiwavelength analysis of EELGs in miniJPAS]{Multiwavelength exploration of Extreme Emission Line Galaxies detected in miniJPAS survey} 

\author[I. Breda et al.]{Iris Breda,$^{1,2}$\thanks{E-mail: iris.breda@univie.ac.at}
Stergios Amarantidis,$^{3}$
Jos\'e M. Vilchez,$^{2}$
Enrique P\'erez-Montero,$^{2}$
Carolina Kehrig,$^{2}$
\newauthor Jorge Iglesias-P\'aramo,$^{2}$
Antonio Arroyo-Polonio,$^{2}$
Juan A. Fern\'andez-Ontiveros,$^{4}$
\newauthor Rosa M. Gonz\'{a}lez Delgado,$^{2}$
Luis A. D\'iaz-Garc\'ia,$^{2}$
Raul Abramo,$^{5}$
Jailson Alcaniz,$^{6}$
Narciso Ben\'itez,$^{2}$
\newauthor Silvia Bonoli,$^{4,7,8}$
Javier A. Cenarro,$^{9}$
David Crist\'obal-Hornillos,$^{4}$
Renato Dupke,$^{6}$
Alessandro Ederoclite,$^{4}$
\newauthor Antonio Hern\'an-Caballero,$^{9}$
Carlos L\'opez-Sanjuan,$^{7}$
Antonio Mar\'in-Franch,$^{9}$
Claudia Mendes de Oliveira,$^{10}$
\newauthor Mariano Moles,$^{4}$
Laerte Sodr\'e,$^{10}$
Keith Taylor,$^{11}$
Jes\'us Varela,$^{9}$
H\'ector V\'azquez-Rami\'o$^{9}$   
\\
$^{1}$Instituto de Astrof\'{i}sica de Andaluc\'{i}a - Glorieta de la Astronom\'{i}a, s/n, 18008 Granada, Spain \\
$^{2}$Department of Astrophysics, University of Vienna, Türkenschanzstraße 17, 1180 Vienna, Austria\\
$^{3}$Instituto de Radioastronom\'ia Milim\'etrica, Av. Divina Pastora 7, N\'ucleo Central E-18012, Granada, Spain \\
$^{4}$Centro de Estudios de F\'isica del Cosmos de Arag\'on, Plaza San Juan,
44001 Teruel, Spain \\
$^{5}$Instituto de F\'isica, Universidade de S\~{a}o Paulo, Rua do Mat\~{a}o 1371, CEP, 05508-090 S\~{a}o Paulo, Brazil \\
$^{6}$Observat\'orio Nacional – MCTI (ON), Rua Gal. Jos\'e Cristino 77,
São Crist\'ov\~{a}o 20921-400, Rio de Janeiro, Brazil \\
$^{7}$Donostia International Physics Centre (DIPC), Paseo Manuel de Lardizabal 4, 20018 Donostia-San Sebasti\'an, Spain \\
$^{8}$IKERBASQUE, Basque Foundation for Science, 48013 Bilbao, Spain  \\
$^{9}$Centro de Estudios de F\'isica del Cosmos de Arag\'on (CEFCA), Unidad Asociada al CSIC, Plaza San Juan 1, 44001 Teruel, Spain  \\
$^{10}$Instituto de Astronomia, Geof\'isica e Ci\^{e}ncias Atmosf\'ericas, Universidade de S\~{a}o Paulo, 05508-090 S\~{a}o Paulo, Brazil  \\
$^{11}$Instruments4, 4121 Pembury Place, La Canada Flintridge, CA 91011, USA 
}

\pubyear{2024}

\begin{document}
\label{firstpage}
\pagerange{\pageref{firstpage}--\pageref{lastpage}}
\maketitle

\begin{abstract}
Extreme Emission Line Galaxies (EELGs) stand as remarkable objects due to their extremely metal poor environment and intense star formation. Considered as local analogues of high-redshift galaxies in the peak of their star-forming activity, they offer insights into conditions prevalent during the early Universe. Assessment of their stellar and gas properties is, therefore, of critical importance, which requires the assembly of a considerable sample, comprehending a broad redshift range. The Javalambre-Physics of the Accelerating Universe Astrophysical Survey (JPAS) plays a significant role in assembling such a sample, encompassing $\sim$8000 $\rm deg^2$ and employing 54 narrow-band optical filters. The present work describes the development and subsequent application of the tools that will be employed in the forthcoming JPAS spectrophotometric data, allowing for the massive and automated characterization of EELGs that are expected to be identified. This fully automated pipeline (requiring only the object coordinates from users) constructs Spectral Energy Distributions (SEDs) by retrieving virtually all the available multi-wavelength photometric data archives, employs SED fitting tools and identifies optical emission lines. It was applied to the sample of extreme line emitters identified in the miniJPAS Survey, and its derived physical properties such as stellar mass and age, coupled with fundamental relations, mirror results obtained through spectral modeling of SDSS spectra. Thorough testing using galaxies with documented photometric measurements across different wavelengths confirmed the pipeline’s accuracy, demonstrating its capability for automated analysis of sources with varying characteristics, spanning brightness, morphology, and redshifts. The modular nature of this pipeline facilitates any addition from the user.
\end{abstract}




\section{Introduction \label{intro}}

Extreme emission line galaxies (EELGs) are a sub-sample of extra-galactic objects which thorough understanding remains elusive. These exceptional astronomical entities, which may seem irrelevant at first sight due to their faint nature and reduced dimensions, might hold the key to our understanding of galaxy formation. 

Typically, EELGs are highly compact objects with vivid blue/green colors, being characterized by very high (frequently reaching three orders of magnitude) equivalent widths (EW) of relatively high-excitation emission lines indicative of intense and ongoing star-formation (SF), such as H$\alpha$, optical [OIII], and [OII] or Ly-$\alpha$ when observed at higher redshifts where they are more frequently detected \citep[e.g.,][]{KunSar86,Erb16}. By presently harbouring the most violent SF events in the Universe, these galaxies provide a window to the past, resembling epochs where global SF was at its zenith \citep[e.g.,][]{Amor15,EPM21}. They produce a substantial amount of photoionising radiation originated from complexes of young, massive stars which, although controversial \citep{LoBa01,Bos18}, is considered by various authors to significantly contribute to the re-ionization of the Universe \citep[e.g.,][]{Sal11,Dre15,Erb16}. In addition, this kind of galaxies are acknowledged as likely being the building blocks of local, more massive galaxies \citep{Dre11}, and are amidst the most metal poor objects observed in the Universe \citep[e.g.,][for a review]{KunOst00,Pap08}. Wolf-Rayet features are often identified in these galaxies \citep[e.g.,][]{Scha99, Amor12}. They might be classified as strong He II emitters \citep[e.g.,][]{Keh18,Fer21} and, depending upon the adopted selection criteria and redshift at which they are observed, as blue compact galaxies or blue compact dwarfs \citep[BCGs/BCDs; e.g.,][]{ThuMar81,LooThu86,KunSar86,Pap96,Cai01,Rev07}, green pea galaxies \citep[GPGs;][]{Car09, Amor15}, blueberries \citep[][]{Yan16} and/or ELdots \citep{Bek15}.

Reviewing the vast complexity of the subject and the broad scientific framework that these exceptional galaxies comprehend, the identification of a statistically meaningful sample across redshift is of uttermost importance. Such will allow to pursue a thorough study of the nature of these objects, and ultimately portray a coherent narrative of their role in the broader context of galaxy formation and evolution. There are several essays attempting for the assembly of a representative sample of EELGs across redshift, such as, e.g., \citealt{Wel11} (70 galaxies in the CANDELS fields at $z \sim$ 1.7), \citealt{Amor10} ($\sim$ 180 galaxies from the 20k zCOSMOS bright survey with 0.11 $\leq$ $z$ $\leq$ 0.93) or \citealt{EPM21} ($\sim$ 2000 galaxies in the SDSS-DR7 with 0 $\leq$ $z$ $\leq$ 0.49) and \citealt{Lum21} (466 EELGs at $z$ < 0.06) using objective prisms to identify strong emission lines. From these studies it was disclosed that EELGs display compact morphologies (with R$_{50}$ $\leq$ 2kpc), and a broad range of stellar masses (6.5 $\leq$ log M$_{\star}$/M$_{\odot}$ $\leq$ 10), with higher masses being mostly found at higher redshifts. Quite importantly, some of these galaxies exhibit pristine environments \citep[e.g.,][]{Izo12,Gri11}, with oxygen abundances ranging from  12 + log O/H = $\sim$7.3 to 8.6, and nitrogen-to-oxygen ratios from $\sim$ 0.01 to 0.1. 
Additional works have provided further insight on this galaxy type. For instance, \citet{Maseda13,Maseda14} has analysed 2 samples of 19 and 22 EELGs detected at 1.3 $\leq$ z $\leq$ 2.3, revealing that these have low stellar masses (10$^{8}$ - 10$^{9}$ M$_{\odot}$) and are experiencing intense starbursts. Additional studies have found similar results, such as the works by \citet{Tran20,Gupta21,Gupta23}, which explores data from the MOSEL survey, covering a sample of strong [OIII] 5007 $\AA$ emitting galaxies at 3 $\leq$ z $\leq$ 4 extracted from the ZFOURGE survey. Furthermore, by analysing the physical properties of 19 EELGs identified in the ZFOURGE survey, \citet{Cohn18} revealed that these galaxies show evidence of a starburst in the most recent 50 Myr, assembling 15\% of their total stellar mass within this short time period, a highly significant value when contrasted with typical star-forming galaxies which in the same time period have formed only 4\% of their total stellar mass. In addition, \citet{Gupta21,Gupta23} have identified signatures of galactic-scale outflows, attributing the pronounced emission lines to interactions and/or mergers. EELGs with even lower stellar masses and higher sSFRs have been identified by \citet{Tang19} at 1.3 $\leq$ z $\leq$ 2.4, ranging from 10$^{7}$ to 10$^{8}$ M$_{\odot}$. In addition, these authors have detected O32 values associated with significant Lyman continuum escape. 

An excellent opportunity to keep expanding the acquired knowledge on this galaxy genus, and to enlarge the number of these galactic specimens, resides on the exploration of the Javalambre-Physics of the Accelerating Universe Astrophysical Survey \citep[JPAS;][]{Bon21}. Such survey is highly suitable for this purpose, considering that it will observe a vast section of the sky ($\sim$8000 deg$^{2}$) by means of 54 narrow band filters in the optical regime with an average full width at half maximum (FWHM) of 145 \AA\ (translated into an R of $\sim$ 60) and an average spatial resolution with a FWHM $<$ 1.5'', spanning from 3780 to 9100 \AA, and 2 additional broad filters extending to the ultra-violet (UV) and near infra-red (NIR). It will yield, when complete, one of the largest repositories of photo-spectra or low (spectral) resolution integral field spectroscopy for all morphological galaxy types from the local Universe up to $z \sim 1$. Moreover, past studies have demonstrated that via this photometric survey one can efficiently identify and characterize emission line galaxies up to $z$ < 0.35 \citep{Gines21,Gines22} and the evolution of both red and blue galaxy populations since $z$ = 1 \citep{Gonz21,Gonz22}. Low and intermediate redshift QSOs should be also easily detected \citep{Que22} as well as extended Lyman alpha QSOs \citep{Rah22}. However, for the correct identification of such elusive objects as EELGs, given their low surface brightness stellar continuum and the similarity with QSOs when observed through the filter scheme of JPAS, our team developed an efficient technique that identifies EELGs candidates by searching for especially strong EW emission lines (EW $\geq 200$ \AA). This method \citep[][hereafter IP22]{IP22} was tested using the data released by the miniJPAS survey (covering 1 square degree of the AEGIS field), having identified 20 EELGs and 11 QSO candidates.

This article describes the characterization of the EELGs $\&$ QSOs identified by the aforementioned tool, outlining the development of the employed fully automated, modular pipeline, and including an informative discussion and comparison of the obtained properties. The strategy here delineated serves as a proof of concept and is expected to be massively employed in the forthcoming JPAS narrow-band data, being well suited to analyse all galaxy types. 
This article is organized as follows: 
Sect. \ref{sample} briefly describes how the EELG and QSO candidates were identified in the AEGIS field through the miniJPAS data, Sect. \ref{meth} outlines each module of the developed pipeline, Sect. \ref{res} presents the obtained results and Sect. \ref{conc} summarizes the main conclusions of the present work.

\section{Identifying the galaxy sample} \label{sample}

The galaxy sample was identified by applying the detection algorithm developed by IP22 to the miniJPAS data. This technique relies on the exploration of the miniJPAS catalogues, thereby constructing and subsequently analysing the photo-spectra of all the listed objects. These catalogues were produced firstly by identifying the numerous detected sources in post-reduced photometric data with \sextractor\ \citep{BerArn16} by adopting two parallel strategies, i.e., dual and single mode \citep[see][for a detailed description of imaging data reduction and post-processing]{Bon21}. In this fashion, several quantities of interest were collected, such as coordinates and fluxes integrated within various apertures. Having as primary objective the correct identification of all EELGs in the photometric data-frames from JPAS, the IP22 detection tool's selection criteria was tailored after thorough examination of real EELGs and QSO spectra, and how such quantitative measurements would translate when visualized through the JPAS filter scheme. This approach resulted in the following procedure for the selection of EELGs candidates in miniJPAS photometric data. Considering the dual catalogue and each of the medium-band JPAS filters from J0400 to J0900, it were selected the sources that fulfil the subsequent criteria:
\\
\hspace*{2.5ex}- For each photometric frame F$_{\lambda}$, only sources with a flux density $\geq$ 10$^{-17}$ erg$\cdot$cm$^{-2}$ $\cdot$ \AA$^{-1}$ $\cdot$ s$^{-1}$ were selected;
\\
\hspace*{2.5ex}- From the previously selected sources, the ones with FLAG $>$ 3 or MASK\_FLAG $>$ 0 were discarded, this way avoiding uncertainties appertaining to instrumental artefacts and false detections;
\\
\hspace*{2.5ex}- To prevent the detection of artificial red objects, sources with f$_{8500}$ / f$_{4300}$ > 1.2 were rejected (where f$_{8500}$ and f$_{4300}$ are the integrated fluxes in the filters J0800 to J0900 and J0378 to J0480, respectively); 
\\
\hspace*{2.5ex}- Only sources detected in both catalogues (single and dual) for each JPAS data frame were considered;
\\
\hspace*{2.5ex}- Only sources with minimum contrast (f$_{\rm n}$ - f$_{\rm c}$)/ f$_{\rm n}$ $\ge$ 0.674 (where f$_{\rm n}$ is the total flux of the source in filter $n$ and f$_{\rm c}$ is the estimated continuum flux density), being comparable to an EW of $\sim$ 300 \AA\ in at least one emission line, were accepted as EELG candidates (if the rest-frame EW $<$ 300 \AA\ the source will be discarded);
\\
\hspace*{2.5ex}- To avoid spurious detections, it were rejected all objects to which the intensity peak in filter $n$ is lower than 5$\sigma$, where $\sigma$ corresponds to the standard deviation of the sky in the given frame $n$;
\\
\hspace*{2.5ex}- The candidates identified by the preceding criteria were ascertained as EELG or QSO by direct comparison of their photo-spectra with a clearly defined sample of SDSS-DR14 spectroscopic data of both SF galaxies and QSOs, as well as with a set of synthetic spectra, after convolution with the medium-band JPAS filters. Additionally, this approach enabled to determine the redshift of the source, by performing a systematic comparison of the  observed photo-spectra with red/blue-shifted SDSS spectra. The latter were shifted from $z$ - 0.05 to $z$ + 0.05 in increments of 0.002, thus covering a wide and continuous redshift range.

This strategy resulted in the successful detection of 31 extreme line emitters, from which 20 were classified as EELGs and as 11 QSO.

\section{Description of the modular pipeline} \label{meth}

In pursuance of attaining complete characterization of the detected sources by extraction of all physical properties that can be estimated from the analysis of photometric data, a fully automated, modular pipeline was developed. This section is dedicated to the description of such pipeline, which was scripted in the \emph{python} programming language  \citep{python}, and besides employing widely used \emph{python} packages such as \emph{numpy}, \emph{scipy}, etc., it exploits the astronomy targeted \emph{python} packages \emph{astropy} \citep{astropy}, \emph{astroquery} \citep{astroquery}, \emph{photutils} \citep{photutils}, \emph{specutils} \citep{specutils} and \emph{sewpy} \footnote{https://github.com/megalut/sewpy} (the \emph{python} implementation of \sextractor). 

As a synopsis, the pipeline herein introduced retrieves and processes imaging data (miniJPAS and virtually all publicly available multi-wavelength data) for the desired source, thus extracting the optical photo-spectra and the spectral energy distribution (SED) in subsequent stages. Thereafter, it de-redshifts and examines the photo-spectra, seeking for emission lines. If any emission line is detected it estimates its properties, in particular central wavelengths, fluxes and EWs. In the following stage, it incorporates all the retrieved data in the analysis and constructs the SED, which will be fitted by means of the SED fitting tools \cigale\ \citep{cigale} and \pros\ \citep[][both delayed-$\tau$ and $\alpha$ models]{prospector}, subsequently estimating restframe SDSS intrinsic colors and magnitudes, fluxes and luminosities in each of the observed passbands. 
As a remark, considering the modular nature of the pipeline, it is straightforward to encompass any additional SED fitting tool that the user may desire. The last module extracts the surface brightness profile (SBP) of the source by utilizing a \emph{python} adaptation of the isophotal annuli (\brem{isan}) surface photometry technique by \citet{P02}. If desired, this module may be used to extract the photo-spectra in each individual \brem{isan}. 

\subsection{Pre-processing of the narrow-band data\label{mod1}}

Although the user may provide additional details such as observed redshift and the presence/absence of active galactic nucleus (AGN), the pipeline merely requires the coordinates of the desired galaxy. The processing begins by locating the input coordinates in the miniJPAS catalogue and downloading the photometric frames (F$_{\lambda}$) and respective weight maps. If the redshift is not provided, it queries the NASA/IPAC extragalactic database\footnote{The NASA/IPAC Extragalactic Database (NED) is operated by the Jet Propulsion Laboratory, California Institute of Technology, under contract with the National Aeronautics and Space Administration.} (NED) for the recorded spectroscopic redshift. In case that such information is not available it will subsequently use the photometric redshift previously extracted from the miniJPAS catalogues. It additionally fetches the E(B-V) value by querying the input coordinates via the \emph{astroquery.irsa\_dust} sub-module. The photometric frames F$_{\lambda}$ will be corrected from the effects of interstellar Galactic dust by following the prescription for the extinction curve by \citet[][herein designated as CCM89]{CCM89}, and adopting R$_{\rm V}$ = 3.1:
\begin{equation}
\rm F_{\lambda}^{\: cor} = F_{\lambda} / 10^{-0.4 \cdot CCM89(\lambda) \cdot A_{V}}
\end{equation}
where $\rm A_{V} = R_{V} \cdot E(B-V)$. 

Supplementary data such as zero point magnitudes and band widths are retrieved from the miniJPAS catalogues. Effective wavelengths ($\lambda_{\rm eff}$) are computed through the filter transmission curves (T$(\lambda)$) in such manner, while considering the nature of the detector:
\begin{equation}
\lambda_{\rm eff} \equiv
\sqrt{\dfrac{\int \rm T(\lambda)\cdot\lambda \: d\lambda}{\int \rm T(\lambda)/\lambda \: d\lambda}},
\end{equation}
for photon counter detector types (namely all JPAS data), and 
\begin{equation}
\lambda_{\rm eff} \equiv 
\sqrt{\dfrac{\int \rm T(\lambda) \: d\lambda}{\int \rm T(\lambda)/\lambda^2 \: d\lambda}},
\end{equation}
for electron counter detector types (such as several of the multi-wavelength data here analysed, namely, 2MASS, WISE, Spitzer and Herschel).

The following procedure is to perform a first square-cut (length of 200'') of all of the extinction corrected miniJPAS frames, and to produce a true-color (RGB) image of the sub-frame. The latter is subsequently fed to \emph{sewpy}. Such will deliver a table listing all the detected objects (which will be considered in later stages, assisting to determine the presence/absence of extra sources within the aperture adopted for the photometric analysis) and the radial extent of our source (R$_{\rm gal}$, i.e., the galactic extent estimated by \sextractor\ in the RGB image), which will be utilized for estimating the aperture radius (R$_{\rm ap}$) in later stages. 

A second square-cut of length of 5$\;$R$_{\rm gal}$ centred in our source is executed in each F$_{\lambda}$ and respective root-mean-square (RMS) map (defined as the square-root of the inverse of the observed frame), and the trimmed imaging data are compressed in one data-cube. Supplementary information that will be useful in later stages of the processing is annexed to the ensuing data-cube, namely the pixel-scale, detector type ('P'/'E' for photon/electron), name and location of the file pertaining the filter transmission curve T($\lambda$), effective lambda $\lambda_{\rm eff}$, band's width, average FWHM of the instrument's point spread function (PSF), zero point magnitude, canonical detection limits in the data units and in mJy, exposure time, AB conversion factor and RMS map.

\subsection{Retrieval of multi-wavelength data \label{mod2}}

By supplying the coordinates, the following module yields nearly all the photometric data that are publicly available for the requested source, and the respective RMS maps (when accessible). It stands to benefit from the Barbara A. Mikulski Archive for Space Telescopes\footnote{The Mikulski Archive for Space Telescopes (MAST) is a NASA funded project to support and provide to the astronomical community a variety of astronomical data archives, with the primary focus on scientifically related data sets in the optical, ultraviolet, and near-infrared parts of the spectrum. MAST is located at the Space Telescope Science Institute (STScI).} (MAST) as a sub-module of the \emph{astroquery}, and IRSA-IPAC \citep{Alex05} API\footnote{The IRSA-IPAC Application Programming Interface may be accessed though https://irsa.ipac.caltech.edu/ibe/sia.html} services, to automatically retrieve the located imaging data. 

In particular, this module seeks:
\\
\hspace*{2.5ex}- the MAST database for the Galaxy Evolution Explorer (GALEX) \emph{Deep Imaging Survey} (DIS) photometric data or, in case it is not available, for GALEX \emph{All-sky Imaging Survey} (AIS) \citep[see][for a review on the mission and main surveys that were carried out]{galex}\footnote{For further information regarding the data reduction pipeline and thorough description of the data provided by the MAST database the reader is addressed to the website: \\
http://galex.stsci.edu/gr6/?page=ddfaq}. The decommissioned space telescope GALEX has collected imaging and spectroscopic data in two ultraviolet bands, Far UV (FUV) 1350 -- 1780 \AA\ and Near UV (NUV) 1770 -- 2730 \AA, providing simultaneous co-aligned FUV and NUV images with spatial resolution (i.e., the FHWM of the PSF) of 4.3 and 5.3'', respectively. The GALEX-DIS covered an area of 80 deg$^2$ with integration times spanning 10ks -- 250ks, with a typical integration time of 30ks, reaching a limiting AB magnitude of the order of 25 mag (3.92e10$^{-4}$, 3.71e10$^{-4}$ mJy for the two respective bands). In contrast, the GALEX-AIS has a typical integration time of 100 s, covering over an area of 26,000 deg$^2$, reaching a depth of m$_{\rm AB}$ of 20 in both bands (0.428 and 0.337 mJy, respectively). GALEX data units of flux are given in counts/s.
\\
\hspace*{2.5ex}- the IRSA-IPAC database obtaining photometric data from the Two Micron All-Sky Survey \citep[2MASS; see][for a description of the pipeline data-reduction and specifics of the data produts]{2mass,2mass2}. The 2MASS All-Sky Data Release covers approximately the entire celestial sphere in the near-infrared, providing imaging data on three bandpasses, namely J (1.235 $\mu$m), H (1.662 $\mu$m,) and K$_{\rm s}$ (2.159 $\mu$m), achieving the canonical sensitivities of $\sim$ 15.8, 15.1 and 14.3 m$_{\rm Vega}$ (0.8, 1.0 and 1.3 mJy), respectively, and a spatial resolution of roughly 3'' for the three bands. 2MASS data units of flux are given in Data Numbers (DN).
\\
\hspace*{2.5ex}- the IRSA-IPAC database for ALL Wide-field Infrared Survey Explorer imaging data (the space-based telescope WISE; see \citealt{wise} and \citealt{allwise} for a detailed description on the pipeline, data-reduction and data specifics). ALLWISE was assembled by combining the entire data from both WISE and NEOWISE \citep{Main11} survey phases. 
It delivers enhanced data products with increased photometric sensitivity and accuracy, and improved astrometric precision as compared to the previously available WISE All-Sky imaging data. The data are comprised of 4 bands, W1, W2, W3 and W4, at the respective central wavelengths of 3.4, 4.6, 12 and 22 $\mu$m, with an average spatial resolution of $\sim$6'' for the first three bandpasses, and $\sim$12'' for W4. The respective canonical sensitivity estimates within 95\% of the retrieved data is, in m$_{\rm Vega}$, of 17.36, 15.97, 11.73 and 8.1 (0.037, 0.079, 0.67 and 5.1 mJy). WISE data units of flux are given in DN.
\\
\hspace*{2.5ex}- the IRSA-IPAC database for Spitzer \citep{Wer04} Enhanced Imaging Products (SEIP \footnote{The reader is address to the "Spitzer Enhanced Imaging Products - Explanatory Supplement", for a detailed description of the SEIP final products, in addition to a thorough outline of the data-reduction procedure. The document can be found at:\\
https://irsa.ipac.caltech.edu/data/SPITZER/Enhanced/SEIP/docs/\\
seip\_explanatory\_supplement\_v3.pdf}), a repository comprising high resolution infrared data acquired during the 5 year Spitzer Space Telescope's cryogenic space mission, across a sky area of over 1500 deg$^2$. From the database it extracts the combined images of all four bands (3.6, 4.5, 5.8, and 8 $\mu$m) of the Infrared Array Camera \citep[IRAC;][]{IRAC}, and the 24 micron band of the Multiband Imaging Photometer for Spitzer \citep[MIPS;][]{MIPS}, with each respective bandpass featuring an average spatial resolution of 1.95, 2.02, 1.88, 1.98 and 6''. The respective sensitivity limits within the AEGIS filed\footnote{Such and more specific information on the AEGIS field can be accessed though https://aegis.ucolick.org/} were estimated as 9e10$^{-4}$, 9e10$^{-4}$, 6.3e10$^{-3}$, 5.8e10$^{-3}$ and 0.03 mJy. Spitzer data units of flux are given in MJy per steradian. Both the science and uncertainty imaging data are converted to units of mJy, by division by the conversion factor (1 steradian = 1 rad$^2$ = 4.25e10$^{10 \; \square}$) and subsequent multiplication by the square of the respective pixel-scale of each passband.
\\
\hspace*{2.5ex}- the IRSA-IPAC database for Herschel\footnote{The "Herschel Product Definition Document", which comprehends a complete guide on Herschel data, including calibration, reduction and specifics on the data products, can be accessed at http://herschel.esac.esa.int/hcss-doc-15.0/print/pdd/pdd.pdf}, in particular the Herschel  Multi-tiered Extragalactic Survey \citep[HerMES;][]{Oli12}, a legacy program that comprehends space acquired imaging data, totaling an area of 380 deg$^2$, at wavelengths between 100 and 500 $\mu$m. It comprises data collected with the Herschel Spectral and Photometric Imaging Receiver \citep[SPIRE;][]{Gri10,Ben13} at 250, 350 and 500 $\mu$m and the Herschel Photodetector Array Camera and Spectrometer \citep[PACS; ][]{Pog10,Bal14} at 100 and 160 $\mu$m. The spatial resolution is of 18.2, 24.9 and 36.3'' for SPIRE bandpasses, with estimated canonical sensitivity limits of 5.8, 6.3 and 6.8 mJy, respectively, and of 6.8 and 10.7'', with a typical sensitivity of 5 and 10 mJy, respectively, for the PACS bandpasses. HerMES-SPIRE data units of flux are given in Jy per steradian, while HerMES-PACS are in Jy. The SPIRE science frames are converted to units of mJy by adopting the aforementioned method and PACS data are reduced by three orders of magnitude.

In the successive step, the herein introduced \emph{python} module trims the images to a reduced area of interest with a length of 5 R$_{\rm gal}$ centered in our source and condenses all the retrieved photometry into one data-cube, which additionally includes all the functional information, as previously described for the miniJPAS data. Lastly, the newly acquired data frames will be corrected from Galactic extinction by performing an equivalent procedure as the one described in the previous sub-sect. (R$_{\lambda}$ values for $10 < \lambda < 46.2$ $\mu$m were retrieved from the IRSA-IPAC "Galactic Dust Reddening and Extinction" service\footnote{https://irsa.ipac.caltech.edu/applications/DUST/} and for $\lambda >$ 46.2 $\mu$m a negligible extinction is assumed).

\subsubsection{Assessment of the pipeline's performance on the automated retrieval of photometric fluxes \label{mod2.1}}

\begin{figure*}
\centering
\rotatebox[origin=c]{0}{\includegraphics[width=1\linewidth]{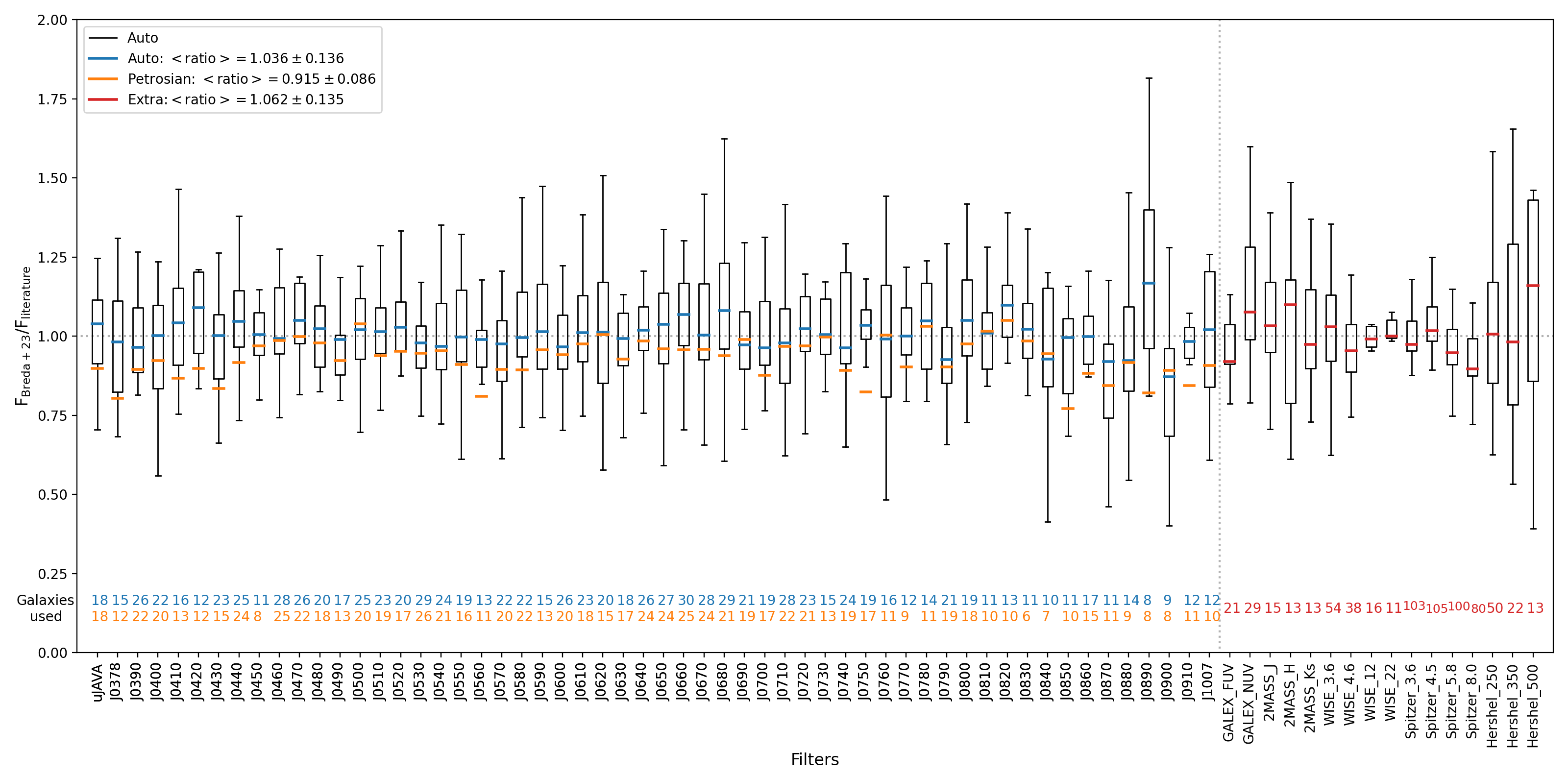}}
\caption{Comparison between the fluxes estimated by this pipeline and the fluxes provided in the literature. The ratios are presented as box-plots containing the 25-75\% percentiles, whilst the upper and lower limits depict the minimum and maximum ratios. The lines inside the box-plots correspond to the median values for the auto, Petrosian and additional bands (blue, orange and red coloured lines), and the legend indicates the mean value of these medians. The coloured values at the bottom of the figure display the number of galaxies that were employed to estimate the box plots for each filter.}
\label{comparison}
\end{figure*}

A series of tests were developed with the goal of evaluating the accuracy of the herein developed pipeline on estimating fluxes through the automatic analysis of photometric frames. Initially, the estimates of the 32 EELGs under study were contrasted with the results from the miniJPAS Public Data Release \citep[MINIJ-PAS-PDR201912;][]{Bon21}. In the interest of adequately compare the two works, both auto and Petrosian fluxes were considered, and to ensure reliability of the data analysis and interpretation, sources with an uncertainty higher that 50\% of the measured flux (i.e., SNR < 2) were discarded. Comparison between the retrieved catalogued fluxes and the ones obtained by the pipeline can be appreciated from Fig.~\ref{comparison}, where the flux ratios are displayed. Inspection of this Fig. reveals that this pipeline comes in excellent agreement with the results from the miniJPAS catalogue (mean values of ratios from all filters of 1.036 and 0.915 for the auto and Petrosian radii, respectively), with the larger differences emerging for the higher-wavelength filters. To note that specially (but not exclusively) for the miniJPAS filters, in the case of no visual detection (i.e., there is no appreciable difference between the flux estimated within the aperture and the surrounding sky), this pipeline is deriving a zero flux. Such can be attributed to the criteria employed by this work in order to exclude possible contaminant sources from the background, thus favouring an accurate photometric analysis of faint galaxies. Additionally, fluxes derived within the Petrosian radius tend to be higher than the ones obtained by using the auto. This disparity can be explained by considering the inherent difference between the Petrosian radius, which in the analysed sample consistently tends to be larger than the auto.

Regarding the additional filters, a similar comparison was conducted. By exploiting the minijpas.xmatch\_allwise catalogue, 10000 randomly selected galaxies were cross-matched (adopting 5~\arcsec as angular separation) with the Hermes-Herschel catalogue \citep{Hermes}, resulting in 1104 sources (including both extended and point-like sources). Adding the 32 galaxies under study, a number of 1135 galaxies were identified and further cross-match with the Spitzer-SEIP catalogue \citep[105 galaxies;][]{Spitz}, 2MASS \citep[74 galaxies;][]{2mass2} and GALEX-AIS \citep[30 galaxies;][]{Bia2011}. All the admitted flux measurements were required to satisfy the criterion of having an uncertainty which is lower than 50\% of its flux measurement, along with being classified as an accurate detection \citep[e.g., photometric quality flag for WISE with a value of B or A; for more details see][]{2mass2}. To note that these criteria and the relatively limited depth of all-sky surveys have led to a reduction in the number of available galaxies in the additional filters.

For all examined filters, the flux ratio between the one retrieved by the pipeline and the reference value was estimated (in the case where solely a magnitude was available from the archival catalogues, the ratio was converted by following $\rm 10^{0.4*|m_{Breda+23}-m_{literature}|}$). Likewise the miniJPAS comparison, the results from this exercise are included in Fig.~\ref{comparison}, with the median displayed with red colour and with a mean ratio value of ~1.062. Although the herein presented is meant to exclusively analyse BCDs observed by JPAS, this exercise further supports its robustness, demonstrating its applicability to both extended and point-like sources, and rendering it qualified for the automated study of both faint and bright galaxies, both in the local and high-z Universe.
   
\subsection{Analysis of the imaging data \label{mod3}}

The present module's first design is to execute aperture photometry on the narrow and medium-band JPAS filters, thuswise extracting the photo-spectra from the imaging data, whereas SED extraction will be performed by the same module in later stages of the processing. Regarding that the construction of both optical photo-spectra and SED share most of the processing stages we decided to compile both proceedings in this sub-sect..


In the interest of selecting an aperture radius R$_{\rm ap}$ that encompasses virtually all of the flux emitted by our source, for miniJPAS data the adopted R$_{\rm ap}$ (pix) is equal to 1.2R$_{\rm gal}$ ('') divided by the pixel scale of each individual frame F$_{\lambda}$. With respect to the multi-wavelength data, given their lower spatial resolution as compared to JPAS, R${\rm ap}$ is set to be twice the size of the FWHM of each individual instrument's PSF, effectively resulting in an aperture of 4 $\times$ FWHM${\lambda}$." Regarding the sky annulus, it is defined as the area between a circle with radius R$_{\rm in}$ = R$_{\rm ap}$ + inc and a second circle with radius R$_{\rm out}$ = R$_{\rm in}$ + inc. The increment inc is adopted to be 10 pixels in all cases except for Herschel data where it is 5, given its large pixel scale. The choice of these values has an empirical foundation, resulting from several trials adopting different values. Two masks centered in the input coordinates are created, mapping the circular area of the aperture (\brem{mask\textsubscript{ap}}) and the sky annulus (\brem{mask\textsubscript{sky}}), which will be subsequently used in the aperture photometry operation. On account of the different pixel-scales of multi-wavelength data, when constructing the SED, the two previously mentioned masks will be created for each multi-wavelength frame.


\begin{figure*}
\centering
\includegraphics[width=1\linewidth]{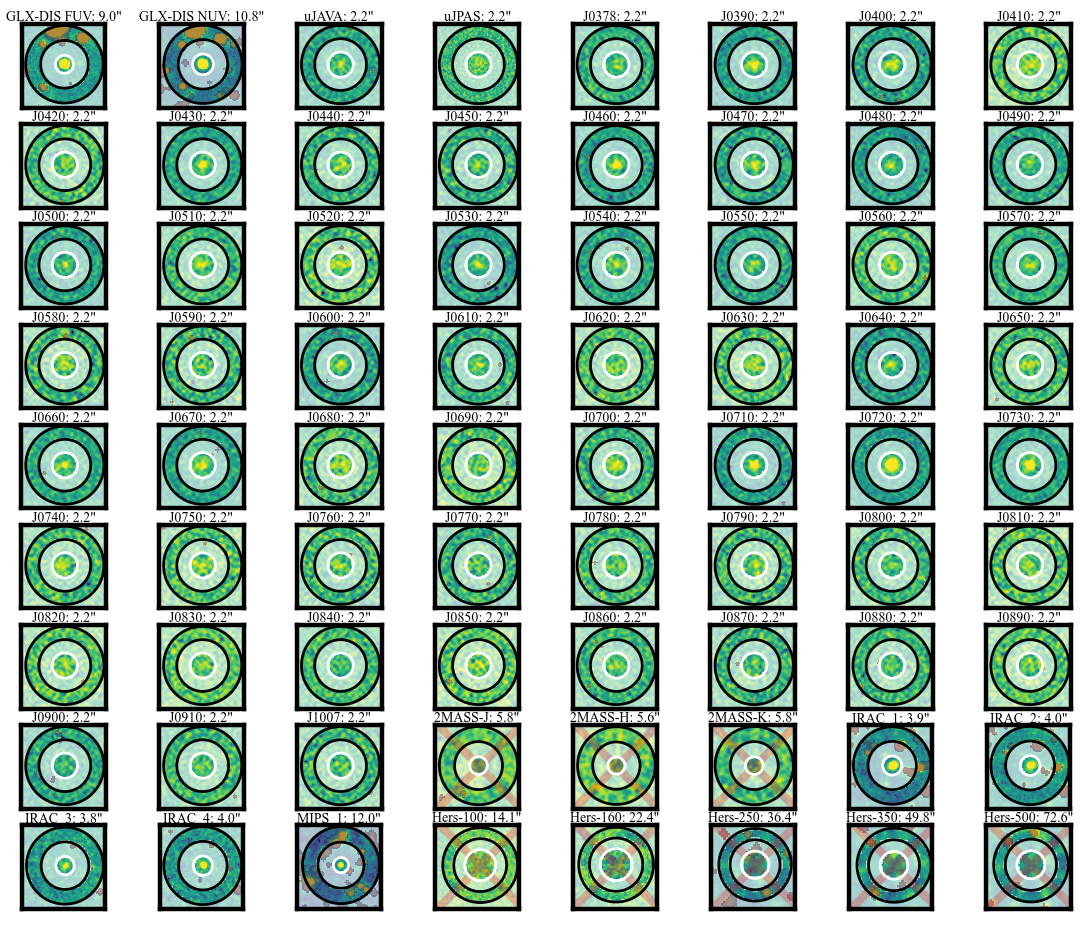}
\caption{Visual output resulting from the module described in sub-section \ref{mod3} for the source located at RA: 214.32208$^{\circ}$, DEC: 52.538199$^{\circ}$ when extracting the SED from multi-wavelength imaging data. It displays all the bandpasses that will be considered in the SED fitting, with the adopted aperture overplotted in white and the sky annulus in black, in addition to the shaded areas which reflect the \brem{maskBS\textsubscript{F\textsubscript{$\lambda$}}}. Moreover, the name of F$_{\lambda}$ and the adopted aperture is noted on top of each frame
and the overlapped red "x" on some of the panels indicates a non-detection.}
\label{all}
\end{figure*}

\subsubsection{Identifying bright sources \label{mod3.1}}

The subsequent phase is to identify and mask all additional bright sources that may exist in each frame, this way mitigating contamination from external sources in the aperture photometry procedure. Aiming for the construction of an algorithm that operates soundly at all circumstances, this is achieved by an elaborated operation which combines several actions:

\begin{itemize}
   \item[a)] After sky modelling and subtraction, for each individual frame create a mask comprising all bright sources (\brem{maskBS\textsubscript{F\textsubscript{$\lambda$}}}), by detecting all excess emission brighter than 3$\sigma$ as compared to the image background. This operation is performed by means of the sigma-clipping method, specifically the sub-module \emph{make\_source\_mask} from the \emph{python} package \emph{photutils}.   
   \item[b)] Identify and extract contours in the mask images, thus obtaining the number, location and dimensions of all the bright sources present in F$_{\lambda}$.
   \item[c)] Obtain \brem{maskBS\textsubscript{F\textsubscript{$\lambda$}-1}} by performing grayscale erosion with a square connectivity equal to one \citep{MIA} in \brem{maskBS\textsubscript{F\textsubscript{$\lambda$}}}, this way minimizing the likelihood of our source to be merged with others (i.e., the same contour to outline our and additional sources).
   \item[d)] Proceed to the identification of contours in \brem{maskBS\textsubscript{F\textsubscript{$\lambda$}-1}} and compare it with the same obtained for the previous set of masks, prior to erosion. If the number of contours have decreased, re-apply the greyscale erosion morphological operation to the original set of masks while adopting a structure of 2$\times$2, thus avoiding neglecting sources with length in one of the spatial dimensions equal to 1 pixel.
   \item[e)] To avoid any risk of neglecting small-scale sources within the aperture radius, the algorithm compares \brem{maskBS\textsubscript{F\textsubscript{$\lambda$}}} with  \brem{maskBS\textsubscript{F\textsubscript{$\lambda$}-1}}, seeking for possible contours with area of 1 pixel that were extinct after the erosion operation. If it detects any it re-places it in \brem{maskBS\textsubscript{F\textsubscript{$\lambda$}-1}}.
   \item[f)] Detect the contour corresponding to our source by identifying the one that is closer to the input coordinates and remove it from \brem{maskBS\textsubscript{F\textsubscript{$\lambda$}-1}}.
   \item[g)] Perform binary dilation twice \citep{MIA} in \brem{maskBS\textsubscript{F\textsubscript{$\lambda$}-1}}, expanding the size of each detected source, thereby obtaining the final mask of bright sources, \brem{maskBS\textsubscript{F\textsubscript{$\lambda$}}}.
   \item[h)] Exclusively for multi-wavelength data, having in mind the substantially broad PSF of several of the instruments here considered, it might be impossible to resolve our source if there is an external bright source in its vicinity. Towards the identification of possible nearby sources, it will estimate the distance between our source and every detected contour. Additionally, it will verify the distances of all objects listed in the \sextractor\ table. To redress this problem, if any of the estimated distances are lower than the PSF of the instrument, the error that will be fed to the SED fitting tools in later stages will be significantly enhanced, assumed to be 50\% of the measured flux within the aperture.
\end{itemize}

\subsubsection{Aperture photometry}\label{mod3.2}

At this stage, all requirements are fulfilled to commence the aperture photometry operation, which is conducted in the following manner: 

\begin{itemize}
   \item[a)] Seclude and extract the flux of our source within the aperture  (and respective sky contribution) firstly by multiplying each F$_{\lambda}$ by the logical NOT (!) of \brem{maskBS\textsubscript{F\textsubscript{$\lambda$}}} (thuswise ensuring that all bright sources within the imaging frame render 0) obtaining F$_{\rm ! maskBS}$, followed by multiplication with \brem{mask\textsubscript{ap}}. The total flux (f$_{\rm ap}$) and number of pixels (n$_{\rm ap}$) within the aperture are assessed by summing the flux of the image resulting from the previous operation, and by accounting for every non-zero, non-nan pixel within this area, respectively. Error estimation ($\sigma_{\rm ap}$) is performed by exploiting the \emph{python} sub-modules \emph{calc\_total\_error} and \emph{aperture\_photometry} from \emph{photutils}, using the individual RMS and the previously compiled exposure times. The number of valid pixels within the aperture (i.e., non-zero and non-nan) is estimated. If lower than 30\% of the total number of pixels within the aperture, this particular F$_{\lambda}$ will not be considered in later stages of the processing.
   \item[b)] Multiply each F$_{\rm ! maskBS}$ by the mask annulus \brem{mask\textsubscript{sky}}. The average flux within the annulus ($\mu_{\rm sky}$) is estimated by summing the image flux resulting from the preceding operation and subsequent division by the number of non-zero pixels within this area. In addition, the mode (Mo$_{\rm sky}$) and the standard deviation of the sky ($\sigma_{\rm sky}$) are estimated.
   \item[c)] The sky flux within the aperture area (f$_{\rm sky}$) is assessed by multiplication of $\mu_{\rm sky}$ by n$_{\rm ap}$. Finally, the corrected flux radiated by our source is given by f$_{\rm gal}$ = f$_{\rm ap}$ - f$_{\rm sky}$.
\end{itemize}

Once the SED is extracted, as a control point, the module offers visual aid as illustrated by Fig. \ref{all} for an example source (galaxy located at RA: 214.32208$^{\circ}$, DEC: 52.538199$^{\circ}$). It is exhibited the various bandpasses that will be considered for this source in the SED fitting procedure, with the white circle depicting the adopted aperture whereas the black annulus indicates the area of the sky used for sky statistics. It is additionally depicted each \brem{maskBS\textsubscript{F\textsubscript{$\lambda$}}} overlapping external bright sources. On top of each frame it is inscribed the name of F$_{\lambda}$ and the adopted aperture. The red "x" indicates a non-detection, i.e., an upper limit for the SED fitting tools.

\subsubsection{Identifying non-detections \label{mod3.3}}

When considering the multi-wavelength data and the subsequent SED fitting operation, it is crucial to devise a strategy for the non-supervised, accurate identification of the presence/absence of emission in the current frame. Intending to determine the criteria that are well-suited for this task, it were drafted a set of trials adopting different conditions that were later evaluated by visual inspection of an exhaustive number of individual frames. This experiment resulted in the choice of the following criteria, used for the determination of a positive detection:

\begin{itemize}
   \item[c1)] the circle's area, encompassing pixels brighter than Mo${\rm sky}$ + 3$\sigma{\rm sky}$ within the aperture, should be greater than or equal to the area of a circle equivalent to the FWHM of the instrument's PSF. In this manner we assure that the detection (herein defined as 3$\sigma_{\rm sky}$ above the sky level) can be resolved by the observing telescope.


    \item[c2)] the fraction of the number of pixels within F$_{\rm ap}$ which are brighter than Mo$_{\rm sky}$ + 3$\sigma_{\rm sky}$ must be higher than the same within f$_{\rm sky}$, assuring that the percentage of pixels that are at least 3$\sigma_{\rm sky}$ above the sky level is superior within the aperture, as compared to the same within the sky annulus.
\end{itemize}

When the algorithm adjudicates a non-detection, the observed flux will be assigned to the SED fitting tools as an upper limit. 

The pipeline continues by performing aperture corrections to all multi-wavelength data whenever deemed necessary (i.e., when R$_{\rm ap}$ is lower enough not to encompass at least 95\% of the source's luminosity, as given by \sextractor), following the recipes provided by each instrument's handbook (see sub-sect. \ref{mod2} for helpful references). Error estimation for the multi-wavelength data that do not provide RMS maps, such as 2MASS, is executed by adopting the prescriptions described in \citet{2mass2}. Regarding Herschel data, as indicated by \citet{Bal14} and \citet{Ben13}, respectively, the flux error is of 5\% in all three PACS filters and 5.5\% in all SPIRE filters.

\subsubsection{Conversion of the data units to physical units of flux \label{mod3.4}}

Subsequently, for each F$_{\lambda}$ that is not in units of mJy, the measured emission f$_{\rm gal}$ is converted to physical units of flux, namely, flux density in f$_{\lambda}$ (erg$\cdot$cm$^{-2}$ $\cdot$ \AA$^{-1}$ $\cdot$ s$^{-1}$) and f$_{\nu}$ (erg$\cdot$cm$^{-2}$ $\cdot$ Hz$\cdot$ s$^{-1}$), mJy and m$_{\rm AB}$. This is achieved by following the prescription:

\begin{equation}
\begin{split}
\rm m_{AB} &= \rm -2.5 \; log_{10}(f_{gal}) + zpt + AB - cF \\
\rm \sigma_{AB} &= (2.5 / \rm ln(10)) \cdot (\sigma_{\rm gal} / \rm f_{gal}) \\
\rm f_{\lambda} &= \rm 10^{(m_{AB} \; + \; 48.60)/-2.5} \cdot (c / \lambda_{\rm eff}^{2}) \\
\rm \sigma_{\lambda} &= \rm f_{\lambda} - \rm 10^{(m_{AB} + \; \sigma_{AB} \; + \; 48.60)/-2.5} \cdot (c / \lambda_{\rm eff}^{2}) \\
\rm f_{\nu} &= \rm 10^{(m_{AB} \; + \; 48.60)/-2.5} \\
\rm \sigma_{\nu} &= \rm f_{\nu} - 10^{(m_{AB} + \; \sigma_{AB} \; + \; 48.60)/-2.5} \\
\end{split}
\end{equation}

where, for each F$_{\lambda}$, zpt is the zero-point magnitude, AB is the AB correction factor between the Vega magnitude system, cF is the aperture correction factor and c is the speed of light in units of \AA/s.

For F$_{\lambda}$ that are in units of mJy, such as all passbands from Spitzer and Herschel, the reverse operations are applied, thus retrieving the flux in the remaining units. Both the optical photo-spectra and the SED in all the aforementioned flux units are stored in two fits files. 

\begin{figure*}
\centering
\includegraphics[width=1\linewidth]{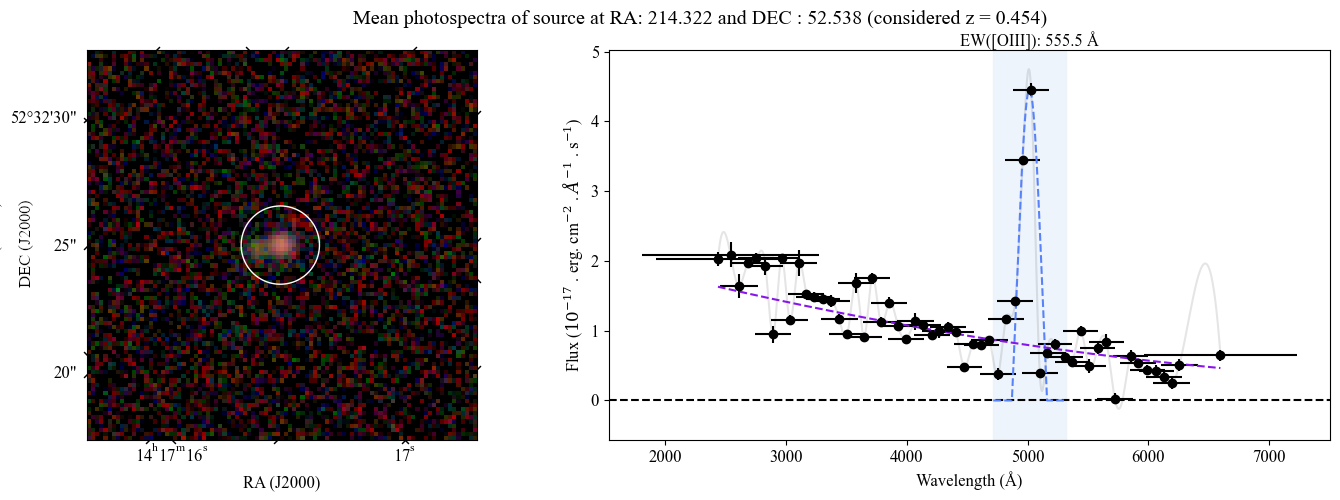}
\caption{Visual output resulting from the module outlined in sub-section \ref{mod3.5}. In the left-hand-side it displays the RGB image of the source with the adopted aperture overplotted in white, and in the right-hand-side it exhibits the rest-frame optical photo-spectra obtained for the source located at RA: 214.32208$^{\circ}$, DEC: 52.538199$^{\circ}$ (and its respective spline-interpolation in grey), along with the [OIII] emission line outlined by the light-blue dashed line. The determined continuum is depicted by the purple dashed line. The shaded light-blue area highlights the region where the emission line was identified.}
\label{mean_spec}
\end{figure*}

\subsubsection{Detecting and assessing emission lines\label{mod3.5}}

After de-redshifting the extracted optical photo-spectra, the subsequent step consists in analysing the latter, seeking for emission lines. The procedure begins by executing the sub-module \emph{find\_lines\_threshold} from the \emph{python} package, \emph{specutils}, which operates by identifying deviations larger than the spectrum's assessed uncertainty. This module was designed for the analysis of spectral data where emission lines are substantially better resolved, with its detection being therefore straightforward. For the case of low resolution spectra such as the one obtained from miniJPAS imaging data, it is frequent for the aforementioned routine to erroneously identify emission lines. To overcome this issue, the list of detections proceeding from this \emph{python} routine will be examined, being contrasted with a record of all the emission lines most commonly observed in star-forming systems, namely, Ly-$\alpha$, CIV, [OII], MgII, H$\beta$, [OIII], H$\alpha$, [NII] \& [SII]. Provided that there are entries to which the observed wavelength differs less than 75 \AA\ from the central wavelength of any of the well-known emission lines, it is considered a true detection.

Modelling of the detected emission lines is performed by means of the 
sub-module \emph{RickerWavelet1D} from \emph{astropy} (after empirically determining that, in most cases, the latter provided better results as compared to a Gaussian fit), thus obtaining the mean (central wavelength) and standard deviation (broadness) for each emission line that was previously detected. The continuum in the vicinity of the emission line is estimated by fitting a second degree polynomial to the photo-spectra, deprived of all deviant points (defined as being lower/higher than the average flux minus/plus its estimated standard deviation). Fluxes and EWs are determined by means of the \emph{specutils} sub-modules \emph{line\_flux} and \emph{equivalent\_width}, respectively, applied in the spectral region between the previously estimated central wavelength of each detected emission line minus/plus its standard deviation. After conducting several tests to establish the threshold of detectability, the empirical limit of 25 \AA\ was selected as the minimum acceptable equivalent width (EW) (note that the narrow bands from JPAS partly overlap, hindering any exact assessment of fluxes and EWs of emission lines). If the detected emission line is listed in the \cigale\ default filters, its integrated flux will be taken into account in the SED fitting operation (the poor precision of the measured fluxes should not obstruct the SED fitting procedure, considering that \cigale\ does not require exact measurements but merely their order of magnitude). An illustrative example is given by Fig. \ref{mean_spec}, which displays the visual output generated by this module for one of the processed galaxies. It is displayed the RGB image in the left-hand-side with the adopted aperture overplotted in white, and the restframe photo-spectra in the right-hand-side, where it is highlighted the detected emission line (most probably resulting from the sum of [OIII]$_{4959}$ and [OIII]$_{5007}$), with an estimated EW of 592 \AA). The light-blue dashed line depicts the fit to the emission line and the purple dashed line the estimated continuum utilized for the determination of the line EW.

\subsubsection{Executing \cigale\label{mod3.6}}

At this stage, the present module initiates the execution of an additional module which addresses the creation of the \cigale\ input files and subsequent execution. For the sake of completeness, we provide a brief description of the necessary files: the \cigale\ data file must contain one row with the galaxy ID, adopted redshift and the flux values in units of mJy. The model configuration file (\emph{pcigale.ini}) will define \cigale's context, encoding the characteristics of the data (namely the passbands to which the values in the data file correspond) and of the models that will be generated to fit the observational data. By default it is assumed a	 Salpeter IMF \citep{Sal55} and a delayed star-formation history (SFH) with optional exponential burst. To note that originally \cigale\ does not have the JPAS filter scheme installed. This operation must be previously performed independently \citep[see][for detailed information on \cigale\ and how to prepare and execute it]{cigale}.

\begin{figure}
\centering
\includegraphics[width=1\linewidth]{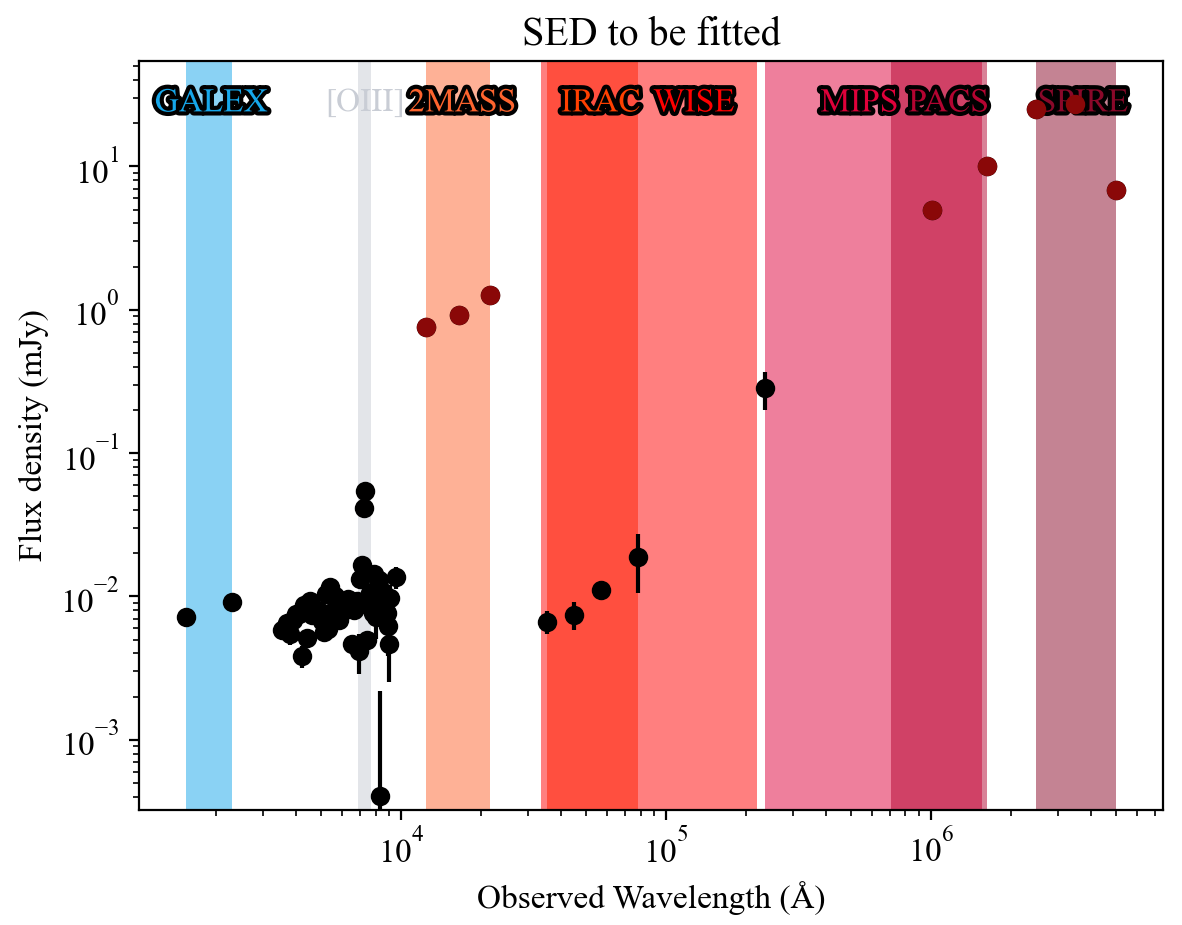}
\caption{Illustration of the observed SED prior of fitting for the sample galaxy displayed in previous figures, with the x axis in logarithmic units and the y axis in units of mJy. The spectral regions of the different passbands are shown, with colors ranging blue to red, and the filed points are color coded according to a detection (black) or non-detection (dark-red). It is additionally highlighted in grey the region where the emission line was detected.}
\label{to_fit}
\end{figure}

As a control point, an additional illustration is produced (Fig. \ref{to_fit}), displaying the observed SED to be fitted by \cigale\ and \pros, subsequently. It indicates the spectral regions of the various pass-bands, color coded from blue to red by increasing wavelength, and the emission lines that were detected in grey. The observational points are displayed in black if considered detections and dark-red if deemed non-detections by the technique elaborated in sub-sect. \ref{mod3.3}.

\begin{figure*}
\centering
\includegraphics[width=1\linewidth]{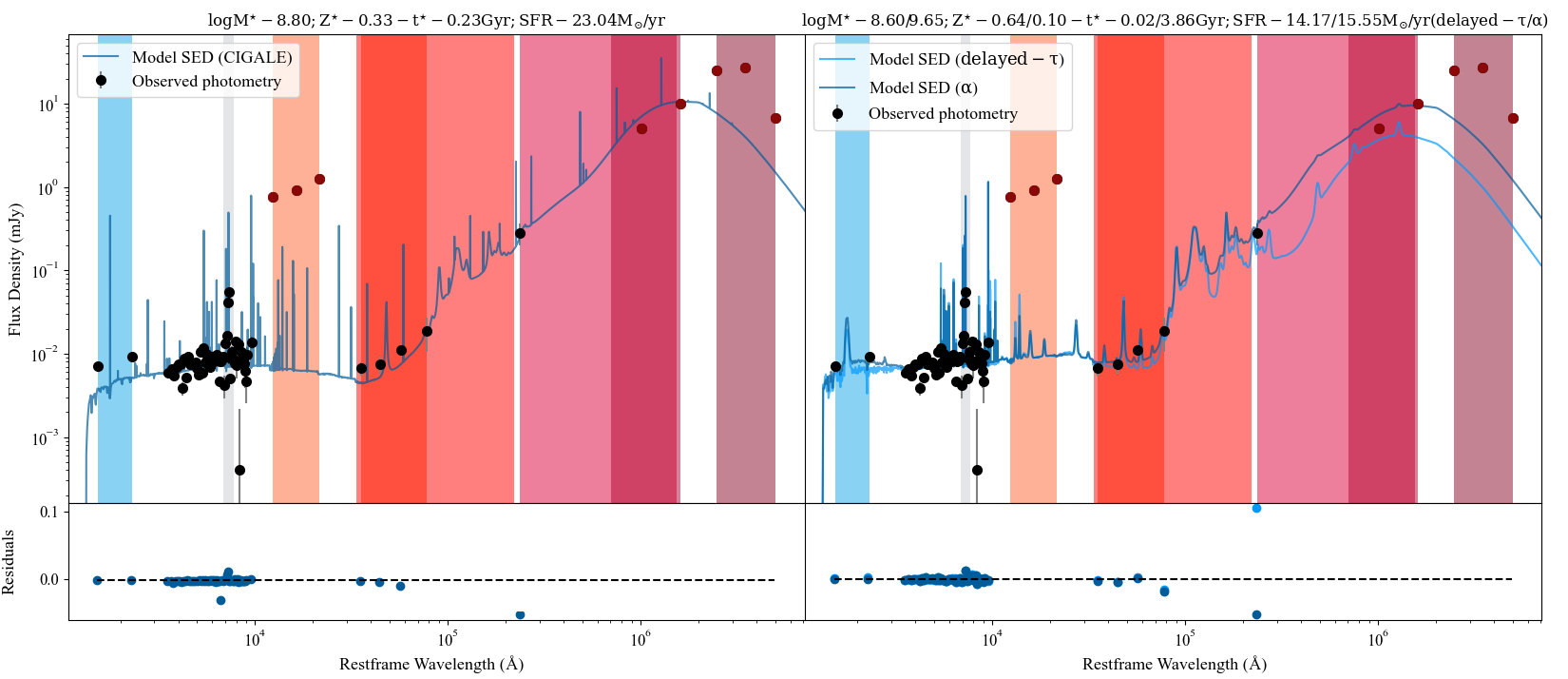}
\caption{Visual output illustrating the SED fitting results by \cigale\ (left-hand side) and \pros\ (right-hand side) for the same exemplar. Both the flux density (in units of 1e$^{-17}$ erg/cm$^{2} \cdot$ s $\cdot$ \AA) and the restframe wavelength (\AA) are displayed in a logarithmic scale. The best-fitting SED models are represented by solid blue lines, the observational points by the black points and the upper limits by the dark red points. The residuals are shown in the low panel and some of the derived physical properties are shown in the image title.
}
\label{sed-fit}
\end{figure*}

Once \cigale's operation is complete, the previously introduced module \ref{mod3} generates a graphical output of the resulting fit, additionally displaying some of the obtained physical properties, as apparent from inspection of the left-hand side of Fig. \ref{sed-fit}. The latter displays the best-fit SED model retrieved by \cigale\ and the observational points and the upper limits in black and red, respectively.


\subsubsection{Executing \pros\label{mod3.6b}}

The module accountable for the execution of \pros\ starts by assembling the two models for inference, implementing a parametric (delayed-$\tau$) and a non-parametric \citep[$\alpha$, ][]{Lej17} star-formation history (SFH). The model referring to the parametric SFH was constructed by selecting the default parametric SFH from \pros\ template library, with a log uniform mass prior, and adding the models for both dust and nebular emission (see \pros\ user guide for a comprehensive description of the provided models). With regard to \pros-$\alpha$, it includes dust attenuation and re-radiation, a flexible attenuation curve, nebular emission, stellar metallicity, and a non-parametric SFH with 6-components. The simple stellar population (SSP) library is produced according to the adopted SFH. In addition, nebular emission and dust (and AGN, if the object was previously classified as QSO) contribution are included. The redshift is fixed and it is adopted a Salpeter IMF. The dictionary encompassing the observational data is assembled, which comprises the list of the utilized passbands, the evaluated fluxes and respective uncertainties in each filter in maggies units, and the effective wavelength for each of the filters. The subsequent step is to optimize the parameters of the calibration functions for each generated model. The posterior probability distribution is estimated by sampling, either through an affine-transform-invariant version of Metropolis-Hastings MCMC ($emcee$) or nested sampling ($dynesty$).
Hereupon, the SED is fitted and the results are stored. Similarly as the \cigale\ module, a graphical output is produced, summarizing the obtained physical properties for both models (such as present-day stellar mass (M$_{\star}$), mass-weighted mean stellar age (t$_{\rm M}$) and metallicity (Z$_{\rm M}$), and SFRs), and displaying the resulting best-fitting models, as exemplified by the right-hand side of Fig. \ref{sed-fit}.

\begin{figure*}
\centering
\includegraphics[width=1\linewidth]{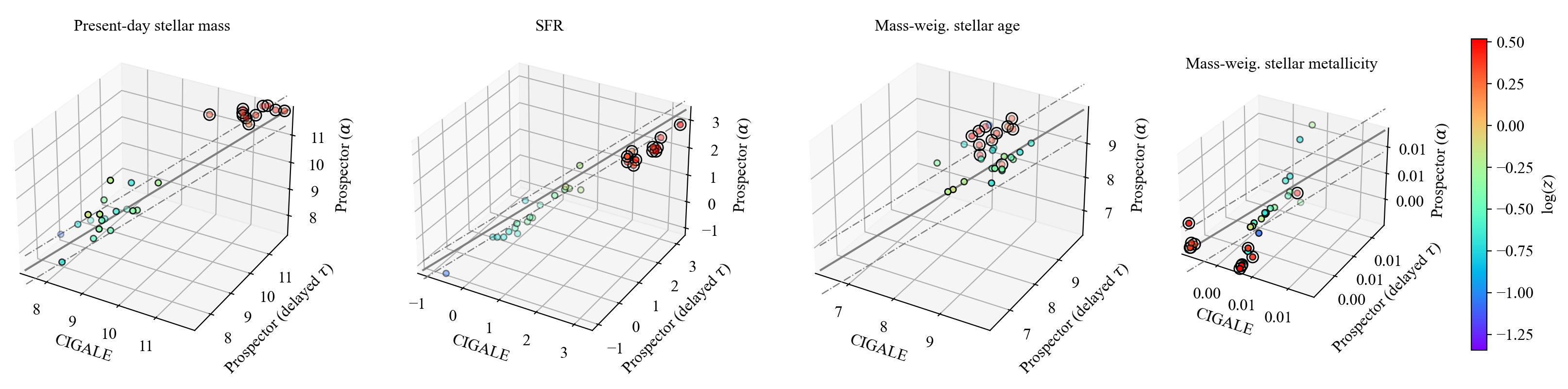}
\caption{Visual representation of the estimates of the present-day stellar mass, SFR, mass-weighted stellar age and metallicity (from left to right) for the three models, for the sample galaxies. Datapoints (EELGs and QSOs) are color coded according to the logarithm of their redshift and QSOs are encircled within larger, empty circles. The solid line represents unity and the dashed lines the average dispersion (1$\sigma$).}
\label{diffs}
\end{figure*}

\subsubsection{Estimate restframe SDSS colors and magnitudes, fluxes and luminosities for each bandpass \label{mod3.7}}

Subsequently, restframe magnitudes, fluxes and luminosities, along with intrinsic colors and apparent and absolute magnitudes in the SDSS filters $i$, $g$ and $r$ are estimated. Such is accomplished by convolving each of the filter transmission curves T$(\lambda)$ with the restframe SED, as subsequently clarified.

Primarily, the restframe best-fitting SED is interpolated to the step of each T$(\lambda)$, followed by its normalization, so that the sum of its flux density is equal to one, resulting in sp$(\lambda)$. Subsequently, to obtain the flux in each bandpass f$\rm _{F_{\lambda}}$ in units of f$_{\lambda}$ (erg$\cdot$cm$^{-2}$ $\cdot$ \AA$^{-1}$ $\cdot$ s$^{-1}$), it is applied the successive equation:

\begin{equation}
\rm f_{\rm F_{\lambda}} = \dfrac{\int \rm sp(\lambda) \cdot \rm T(\lambda)\cdot\lambda \: d\lambda}{\int \rm sp(\lambda) \cdot \rm T(\lambda) \: d\lambda},
\end{equation}

\begin{figure*}
\centering
\includegraphics[width=1\linewidth]{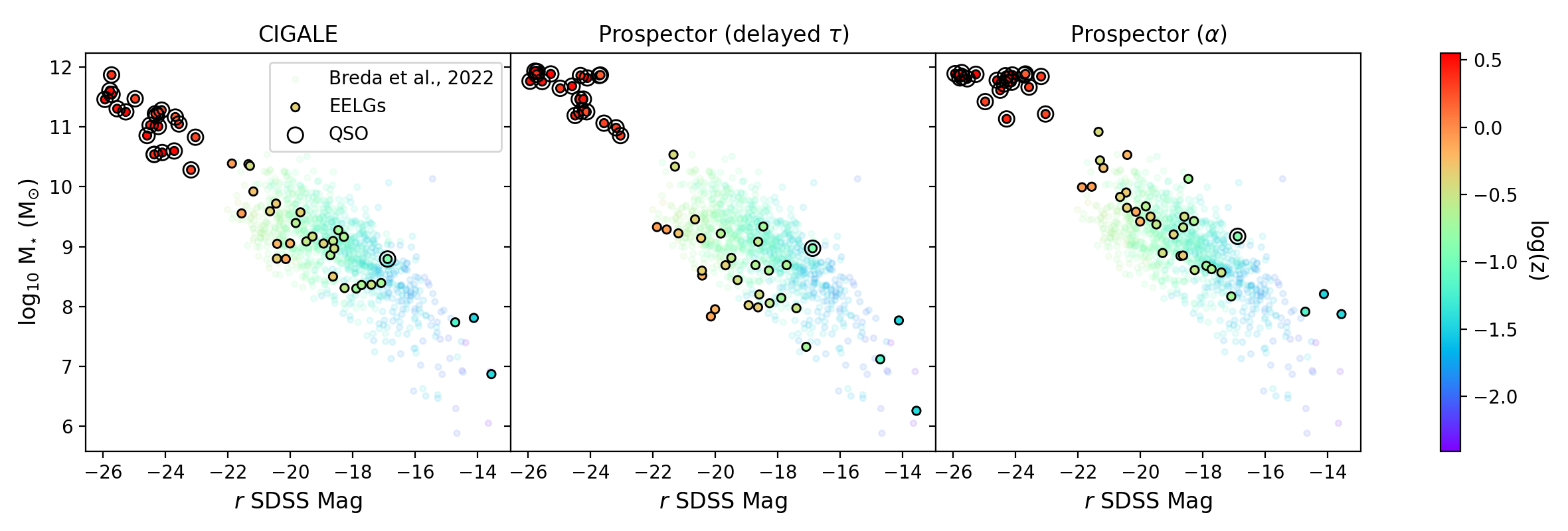}
	\caption{Absolute magnitude in the $r$ SDSS optical band versus the estimated present-day stellar mass M$_{\star}$ as obtained by \cigale\ (left-hand side column), \pros\ delayed-$\tau$ (central column) and \pros-$\alpha$ (right-hand side column). Analogously to Fig.~\ref{diffs}, data-points are color coded according to the logarithm of their redshift and QSOs are encircled within larger, empty circles. The transparent points represent the reference EELG sample.}
\label{mags}
\end{figure*}

Conversion to f$_{\nu}$ units (erg$\cdot$cm$^{-2}$ $\cdot$ Hz$\cdot$ s$^{-1}$) is performed by multiplying f$_{\rm F_{\lambda}}$ by $\lambda_{\rm eff}^2$ / c $\times$ 1e10$^{23}$, obtaining f$\rm \nu_{F_{\lambda}}$, which in turn is converted to apparent and absolute magnitudes\footnote{SDSS magnitudes are obtained by processing the unattenuated best-fitting SED} (m/M) through:

\begin{equation}
\begin{split}
\rm m &= \rm -2.5 \; log_{10}(f\nu_{F_{\lambda}}/3631), \\
\rm M &= \rm m - 25 + 5 \; log_{10}(D),
\end{split}
\end{equation}

where D is the galaxy distance in units of Mpc. Restframe SDSS colors are estimated through:

\begin{equation}
\begin{split}
(g - i) &= \rm m_{g} -  m_{i},\\
(g - r) &= \rm m_{g} -  m_{r},\\
(r - i) &= \rm m_{r} -  m_{i}.
\end{split}
\end{equation}

Finally, restframe luminosities within each bandpass L$_{\rm F_{\lambda}}$ are estimated by:

\begin{equation}
\rm log(L_{\rm F_{\lambda}}) = \rm f_{\rm F_{\lambda}} \cdot 4\pi R^2 ,
\end{equation}

where R is equal to D in units of cm.


\section{Exploration of the physical properties of the miniJPAS EELGs}\label{res}

The coordinates of the 31 extreme line emitters identified by the IP22 algorithm (20 EELGs and 11 QSOs) in the miniJPAS data were provided to the pipeline described in the previous section
. In this fashion, their multi-wavelength photometry was retrieved and analysed by both \cigale\ and \pros\ (delayed-$\tau$ and $\alpha$ SFH models), thus obtaining stellar mass, mass-weighted stellar age estimates, SFRs, AGN fractions, etc.. In addition, by convolving the resulting SED with the filter transmission curves of each instrument, we retrieved absolute magnitudes and colors. The collected results were contrasted with the physical properties of a reference sample of $\sim$ 500 EELGs, as obtained by processing their SDSS optical spectra with \fado\ \citep{GomPap17}, as provided by \citet{Bre22}.

Certain galactic properties assessed by the various SED fitting operations here probed are displayed in Fig. \ref{diffs}, which illustrates, from left to right, the estimates for the stellar mass, SFRs, mass-weighted mean stellar age and metallicity. The solid line represents the equality line and the dashed lines denote the mean dispersion of the differences between the estimates obtained for each model. Inspection of this figure evidences that the stellar mass and the SFRs are fairly consistent (the mean $\sigma$ of the differences between models is of 0.5 dex for the stellar mass estimates and of 0.35 M$_{\odot}$/yr for the SFR), although \pros\ delayed-$\tau$ tends to provide lower stellar mass underestimates. Nevertheless, it is evident that, as expected, QSOs register rather higher differences as compared with the EELGs. Regarding the stellar properties, the mass-weighted mean stellar ages display a higher degree of variation, with an average dispersion of 0.73 dex, whereas the mass-weighted mean stellar metallicities exhibit a mild degree of inconsistency between the three sets of estimates (with their differences displaying an average $\sigma$ of 0.16 Z$_{\odot}$). Such result is also observed in previous works using state-of-the-art spectral synthesis techniques, which are the most reliable method available for the characterization of stellar populations in galaxies \citep[e.g.,][and references within]{Bre22}.

\begin{figure*}
\centering
\includegraphics[width=1\linewidth]{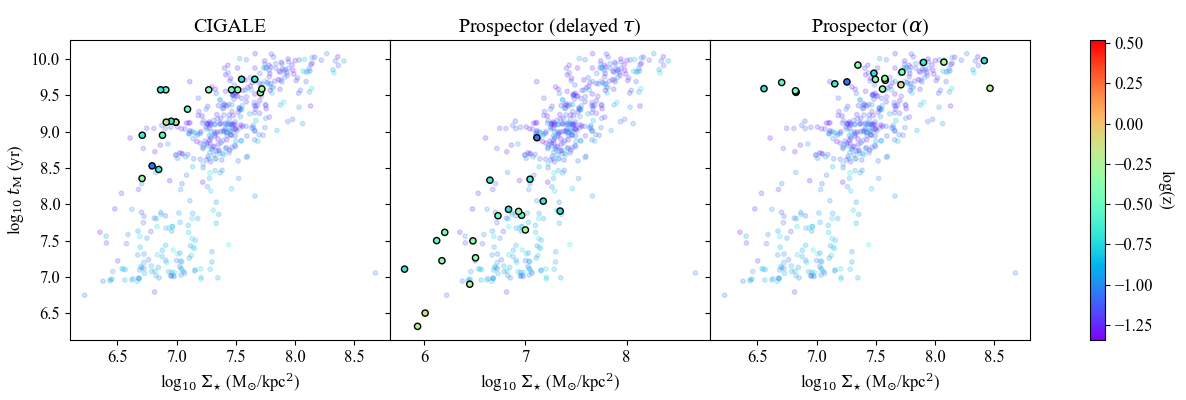}
\caption{Relation between the logarithm of the stellar surface density $\Sigma_{\star}$ and the logarithm of the estimated mass-weighted stellar age t$_{\rm M}$ as obtained for the three SED fitting operations and contrasted with the same for the reference sample.}
\label{age}
\end{figure*}

Figure \ref{mags} illustrates the stellar mass estimates versus the absolute magnitude in the three SDSS bands $g$, $r$ and $i$, as obtained by both SED fitting tools, simultaneously while contrasting with the same for the reference EELG sample. Generally it is observed a sound agreement between the EELGs here in study and the reference sample, with the results by \cigale\ and \pros-$\alpha$ displaying higher coherency and lower levels of dispersion. In contrast, stellar masses recovered by \pros\ delayed-$\tau$ tend to be systematically lower for the respective intrinsic magnitudes, as compared to the reference sample, being in average 0.42 dex lower as compared to the stellar masses recovered by \cigale. These findings are remarkably consistent with previous results -- in \citet{Low20}, the authors compare the derived galaxy properties obtained by SED fitting, by assuming a parametric and a non-parametric model, for a set of $\sim$ 1600 mock galaxies from the SIMBA cosmological simulation \citep{Dave19}, with stellar masses ranging from 4.4$\times$10$\rm^{7}$ to 1.4$\times$10$\rm^{12}$. As illustrated by their Fig.3, these authors report that the estimated stellar masses assuming the \pros\ delayed-$\tau$ model were, on average, approximately $\sim$0.4 dex lower than the true stellar mass. In contrast, the \pros-$\alpha$ model recovers more accurately the true stellar masses within their sample. Regarding the QSOs, as expected, these are located at the high mass vs. high magnitude locus, yet following the same extrapolated trend for the SDSS EELG sample, demonstrating that the stellar mass estimates obtained for the EELGs and QSOs identified in miniJPAS are sound. An additional test was conducted by running the sample exclusively using optical JPAS data. Such resulted in a systematically higher SFR ($\sim$ 0.42 dex, in average), with no significant variation in the remaining parameters. This result is not surprising, considering that the FIR photometry which constrain the dust temperature, is tightly correlated with the level of SFR (the FIR photometric points, or upper limits in most cases, compel the solution to converge to lower dust temperatures, implying lower SFRs).

An additional test was conducted by running the sample exclusively using optical JPAS data. The obtained results remained consistent with those derived from utilizing the complete set of available photometric data, showing slightly higher SFRs.

The stellar surface density was computed by dividing the obtained stellar mass estimates by the projected area (i.e., $\Sigma_{\star}$ = M$_{\star}$ / $\pi$ R$_{\rm gal}^{2}$). The relations between the logarithm of $\Sigma_{\star}$ and the estimated logarithm of the mass-weighted stellar age is shown in Fig. \ref{age} for the three SED fitting operations (from left to right, \cigale, \pros\ delayed-$\tau$ and \pros-$\alpha$). Although there are severe differences between the mass-weighted stellar age estimates as given by the three SED fitting procedures, they all roughly follow the reference correlation. 

\begin{figure*}
\centering
\includegraphics[width=1\linewidth]{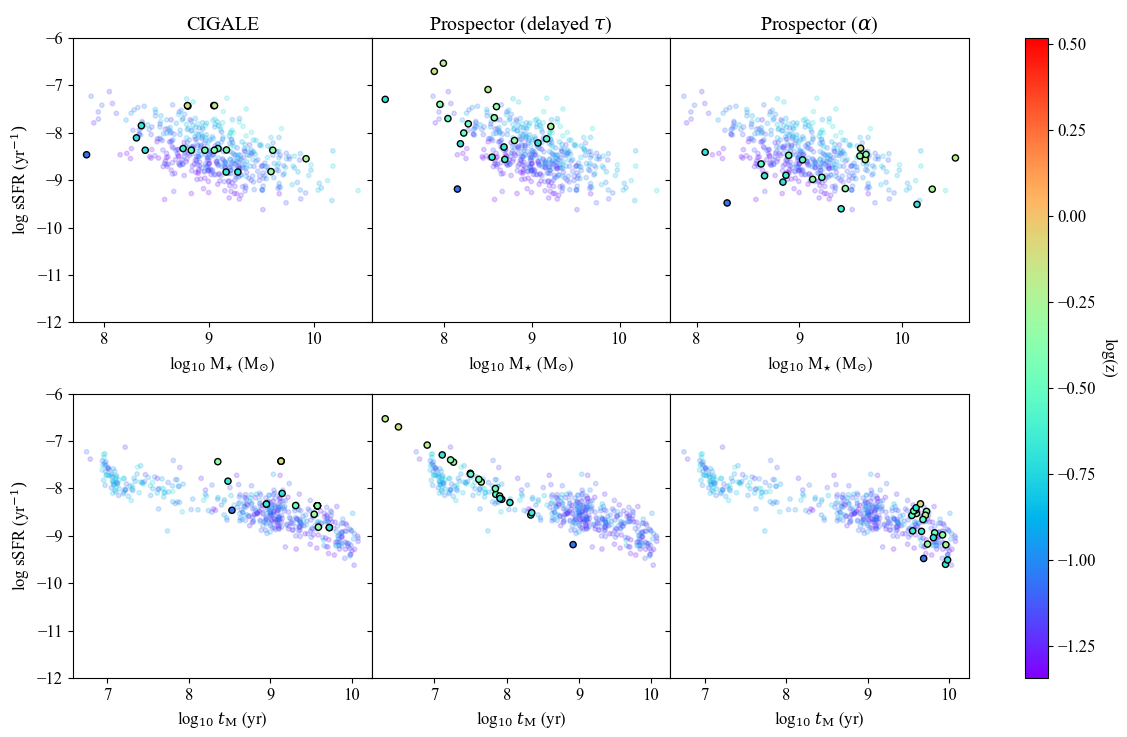}
\caption{The top panels illustrate the logarithm of the present-day stellar mass M$_{\star}$ versus the logarithm of the specific star formation rate sSFR, i.e., the SFR over the estimated stellar mass. The bottom panels display the logarithm of the mass-weighted stellar age t$_{\rm M}$ versus the logarithm of the specific star formation rate. The three panels display the respective results for each of the three SED fitting operations (from left to right, \cigale, \pros\ delayed-$\tau$ and \pros-$\alpha$).}
\label{sSFRs}
\end{figure*}

Finally, the top panels of Fig. \ref{sSFRs} illustrate the logarithm of the present-day stellar mass M$_{\star}$ versus the logarithm of the specific star formation rate sSFR (i.e., the SFR divided by the estimated stellar mass) whereas the bottom panels display the relation between the estimated mass-weighted mean stellar age t$_{\rm M}$ and the logarithm of the sSFR, for the three SED fitting procedures. 




Comparison between the three SED fitting operations indicates that the obtained results are consistent with the reference correlations. Nevertheless, although the stellar mass estimates and sSFRs are fairly coherent between models, the parametric model (\pros\ delayed-$\tau$) tends to underestimate the total stellar mass as compared to the remaining determinations, as previously noted by \citet{Leja19a,Leja19b,Leja20}. Regarding mass-weighted stellar ages and metallicities, these display higher level of variance. As clearly shown by the middle panel of the bottom row of Fig.~\ref{sSFRs}, mean stellar ages recovered by \pros\ delayed-$\tau$ tightly correlate with the sSFR. Such result is expected considering that stellar ages are a direct function of the SFH (i.e., the adopted $\tau$), being analytically computed from the assumed SFH, contrasting with the Dirichlet SFH prior (\pros-$\alpha$) which only weakly couples the SFR to the earlier SFH \citep{Leja17,Leja18}. This outcome reinforces the notion that parametric methods are over-simplistic and incapable of realistically recovering the complex SFHs of galaxies as the EELGs studied herein \citep[e.g.,][]{Low20}. Inspection of this panel additionally reveals an inconsistency between the theoretical prediction and observations, strongly suggesting that an exponentially delayed SFH with a specific $\tau$ is over-simplistic, being unable to appropriately describe the SFH of EELGs, which are expected to be considerably more complex. On the other hand, the mean stellar ages estimated by \pros-$\alpha$ tend to accumulate at the high end of the parameter space. For the lower mass galaxies, the mass-weighted stellar age given by this model appear over-estimated as compared to the reference sample, as reflected by the third panel of Fig. \ref{age}.

Besides the registered inconsistencies, mainly in stellar ages and metallicities, overall the obtained relations
indicate that the developed pipeline produces physically reasonable results and is an appropriate tool for the automatic multi-wavelength analysis and assessment of the main properties of galaxies which lack spectroscopic information. 

 



\section{Summary and conclusions}\label{conc}

The present work describes the development of a fully automated, modular pipeline with the main purpose of the characterization of galaxies with absent spectroscopic information, through a multi-wavelength exploration. The pipeline, scripted in the \emph{python} programming language, fully open-access and offered to the community via the repository GitHub, requires only the coordinates of the desired object. 

It commences by retrieving the available photometric information of the source by seeking in several repositories, namely in the miniJPAS, MAST and IRSA-IPAC databases, collecting photometric data from GALEX, JPAS, 2MASS, WISE, Spitzer and Herschel, and the respective RMS maps, when available. Subsequently, and after masking all bright sources apart from the one under study, it performs aperture photometry in all available passbands, with adapted apertures according to the optical radial extent of the object and the respective FWHM of each instrument. It follows by identifying non-detections, which will be interpreted as upper limits in the SED fitting operations while constructing the observed SED. Additionally, the pipeline detects and assesses emission lines in the optical data, whose integrated fluxes will be used as input in the SED fitting with \cigale.
Finally, it initiates the SED fitting procedures by means of two distinct SED fitting tools (\cigale\ and \pros), for three individual models. 

This strategy, contrary to retrieving data from the available catalogues with fixed apertures, assures consistency and homogeneity throughout the whole set of multi-wavelength data, which is particularly important, specially in light of the faint nature of these objects. The obtained results are in overall agreement with previously known relations derived from detailed spectral synthesis of a sample of $\sim$ 500 EELGs from SDSS, reassuring reliability of the developed method. In addition, the soundness of the retrieved photometry was assessed, revealing solid agreement between the reference values, and demonstrating that the herein developed pipeline is suitable for the automated study of bright and faint sources, point-like or extended, and at local or high-z. 

While EELGs are infrequent in our nearby cosmic neighborhood, their study holds great significance, since these serve as valuable counterparts within our local environment mirroring the characteristics of star-forming galaxies from the early Universe, similar to those newly uncovered by the James Webb Space Telescope \citep[JWST; e.g.,][]{jwst}. This work serves as a prof of concept, demonstrating the reliability of the developed pipeline, paving the way for attaining massive and automated characterization of the EELGs that will be soon detected within the $\sim$ 8000 deg$^{2}$ of the northen sky covered by JPAS.

\section*{Acknowledgements}
I.B., J.V.M., J.I.P., C.K., E.P.M, A.A.P. and R.G.D. acknowledge financial support from the State Agency for Research of the Spanish MCIU through the "Center of Excellence Severo Ochoa" award to the Instituto de Astrofísica de Andalucía (SEV-2017-0709). J.V.M., J.I.P., C.K., and E.P.M. acknowledge financial support from projects Estallidos7 PID2019-107408GB-C44 (Spanish Ministerio de Ciencia e Innovacion), and grant P18-FR-2664 (Junta de Andaluc\'ia). R.G.D. and L.A.D.G. acknowledge financial support from the State Agency for Research of the Spanish MCIU through 'Center of Excellence Severo Ochoa' award to the Instituto de Astrofísica de Andalucía (SEV-2017-0709) and CEX2021-001131-S funded by MCIN/AEI/10.13039/501100011033. L.A.D.G., R.M.G.D., R.G.B., G.M.S., and J.E.R.M. are also grateful for financial support from the projects PID-2019-109067-GB100 and PID2022-141755NB-I00.


Based on observations made with the JST250 telescope and PathFinder camera for the miniJPAS project at the Observatorio Astrof\'{\i}sico de Javalambre (OAJ), in Teruel, owned, managed, and operated by the Centro de Estudios de F\'{\i}sica del  Cosmos de Arag\'on (CEFCA). We acknowledge the OAJ Data Processing and Archiving Unit (UPAD) for reducing and calibrating the OAJ data used in this work.

Funding for OAJ, UPAD, and CEFCA has been provided by the Governments of Spain and Arag\'on through the Fondo de Inversiones de Teruel; the Aragonese Government through the Research Groups E96, E103, E16\_17R, and E16\_20R; the Spanish Ministry of Science, Innovation and Universities (MCIU/AEI/FEDER, UE) with grant PGC2018-097585-B-C21; the Spanish Ministry of Economy and Competitiveness (MINECO/FEDER, UE) under AYA2015-66211-C2-1-P, AYA2015-66211-C2-2, AYA2012-30789, and ICTS-2009-14; and European FEDER funding (FCDD10-4E-867, FCDD13-4E-2685).


\section*{Data Availability}
The data underlying this article are publicly available in the miniJPAS archive, at \url{https://www.j-pas.org/datareleases/minijpas_public_data_release_pdr201912}.




\bsp	
\label{lastpage}

\begin{thebibliography}{}

\bibitem[Alexov et al., 2005]{Alex05} Alexov A., Berriman G.~B., Chiu N.-M., Good J.~C., Jarrett T.~H., Kong M., Laity A.~C., et al., 2005, ASPC, 347, 7
\bibitem[Amor{\'\i}n et al., 2012]{Amor12} Amor{\'\i}n R., P{\'e}rez-Montero E., V{\'\i}lchez J.~M., Papaderos P., 2012, ApJ, 749, 185. doi:10.1088/0004-637X/749/2/185
\bibitem[Amor{\'i}n et al., 2015]{Amor15}Amor{\'\i}n R., P{\'e}rez-Montero E., Contini T., V{\'\i}lchez J.~M., Bolzonella M., Tasca L.~A.~M., Lamareille F., et al., 2015, A\&A, 578, A105. doi:10.1051/0004-6361/201322786
\bibitem[Amor{\'i}n et al., 2010]{Amor10}Amor{\'i}n, R., P{\'e}rez-Montero, E., \& V{\'i}lchez, J. M. 2010, ApJ, 715, L128
\bibitem[Astropy Collaboration et al., 2018]{astropy}Astropy Collaboration, Price-Whelan A.~M., Sip{\H{o}}cz B.~M., G{\"u}nther H.~M., Lim P.~L., Crawford S.~M., Conseil S., et al., 2018, AJ, 156, 123. doi:10.3847/1538-3881/aabc4f
\bibitem[Balog et al., 2014]{Bal14} Balog Z., M{\"u}ller T., Nielbock M., Altieri B., Klaas U., Blommaert J., Linz H., et al., 2014, ExA, 37, 129. doi:10.1007/s10686-013-9352-3
\bibitem[Bekki, 2015]{Bek15}Bekki, K. 2015, MNRAS, 454, L41
\bibitem[Bendo et al., 2013]{Ben13} Bendo G.~J., Griffin M.~J., Bock J.~J., Conversi L., Dowell C.~D., Lim T., Lu N., et al., 2013, MNRAS, 433, 3062. doi:10.1093/mnras/stt948
\bibitem[Bertin \& Arnouts (1996)]{BerArn16}Bertin, E. \& Arnouts, S. 1996, A\&AS, 117, 393.
\bibitem[Bianchi et al., 2011]{Bia2011} Bianchi L., Herald J., Efremova B., Girardi L., Zabot A., Marigo P., Conti A., et al., 2011, Ap\&SS, 335, 161. doi:10.1007/s10509-010-0581-x
\bibitem[Bonoli et al., 2021]{Bon21} Bonoli S., Mar{\'\i}n-Franch A., Varela J., V{\'a}zquez Rami{\'o} H., Abramo L.~R., Cenarro A.~J., Dupke R.~A., et al., 2021, A\&A, 653, A31. doi:10.1051/0004-6361/202038841
\bibitem[Bosch-Ramon, 2018]{Bos18} Bosch-Ramon V., 2018, A\&A, 617, L3. doi:10.1051/0004-6361/201833952
\bibitem[Bradley et al., 2020]{photutils}Larry Bradley L., Sip{\H o}cz B., Robitaille T., Tollerud E., Vin{\'i}cius Z., Deil C., Barbary K., Wilson T. J., Busko I., et al, 2020, Zenodo. doi:10.5281/zenodo.4044744
\bibitem[Breda et al.(2022)]{Bre22} Breda I., Vilchez J.~M., Papaderos P., Cardoso L., Amorin R.~O., Arroyo-Polonio A., Iglesias-P{\'a}ramo J., et al., 2022, arXiv, arXiv:2205.07660
\bibitem[Burgarella, Buat, \& Iglesias-P{\'a}ramo (2005)]{cigale} Burgarella D., Buat V., Iglesias-P{\'a}ramo J., 2005, MNRAS, 360, 1413. doi:10.1111/j.1365-2966.2005.09131.x
\bibitem[Cair{\'o}s et al., 2001]{Cai01}Cair{\'o}s, L. M., V{\'i}lchez, J. M., Gonz{\'a}lez P{\'e}rez, J. N., et al. 2001, ApJS, 133, 321
\bibitem[Cardamone et al., 2009]{Car09}Cardamone, C., Schawinski, K., Sarzi, M., et al. 2009, MNRAS, 399, 1191
\bibitem[Cardelli, Clayton \& Mathis (1989)]{CCM89} Cardelli J.~A., Clayton G.~C., Mathis J.~S., 1989, ApJ, 345, 245. doi:10.1086/167900
\bibitem[Cohn et al., 2018]{Cohn18} Cohn J.~H., Leja J., Tran K.-V.~H., Forrest B., Johnson B.~D., Tillman M., Alcorn L., et al., 2018, ApJ, 869, 141. doi:10.3847/1538-4357/aaed3d
\bibitem[Cutri et al., 2003]{2mass2} Cutri, R. M., et al. 2003, Explanatory Supplement to the 2MASS All Sky DataRelease (Washington: NASA), http://www.ipac.caltech.edu/2mass/releases/allsky/doc/explsup.htm
\bibitem[Cutri et al., 2021]{allwise} Cutri R.~M., Wright E.~L., Conrow T., Fowler J.~W., Eisenhardt P.~R.~M., Grillmair C., Kirkpatrick J.~D., et al., 2013, wise.rept
\bibitem[Dav{\'e} et al., 2019]{Dave19} Dav{\'e} R., Angl{\'e}s-Alc{\'a}zar D., Narayanan D., Li Q., Rafieferantsoa M.~H., Appleby S., 2019, MNRAS, 486, 2827. doi:10.1093/mnras/stz937
\bibitem[Dressler et al., 2011]{Dre11}Dressler, A., Martin, C. L., Henry, A., et al. 2011, ApJ, 740, 71
\bibitem[Dressler et al., 2015]{Dre15}Dressler, A., Henry, A., Martin, C. L., et al. 2015, ApJ, 806, 19
\bibitem[Earl et al., 2020]{specutils} Earl N., Tollerud E., Jones C., Kerzendorf W., Shaileshahuja, D'Avella D., Robitaille T., et al., 2020, Zenodo. doi:10.5281/zenodo.1421356
\bibitem[Erb et al., 2016]{Erb16}Erb, D. K., Pettini, M., Steidel, C. C., et al. 2016, ApJ, 830, 52
\bibitem[Fazio et al., 2004]{IRAC} Fazio G.~G., Hora J.~L., Allen L.~E., Ashby M.~L.~N., Barmby P., Deutsch L.~K., Huang J.-S., et al., 2004, ApJS, 154, 10. doi:10.1086/422843
\bibitem[Fern{\'a}ndez et al., 2021]{Fer21} Fern{\'a}ndez V., Amor{\'\i}n R., P{\'e}rez-Montero E., Papaderos P., Kehrig C., V{\'\i}lchez J.~M., 2021, MNRAS.tmp. doi:10.1093/mnras/stab3150
\bibitem[Ginsburg et al., 2019]{astroquery}Ginsburg A., Sip{\H{o}}cz B.~M., Brasseur C.~E., Cowperthwaite P.~S., Craig M.~W., Deil C., Guillochon J., et al., 2019, AJ, 157, 98. doi:10.3847/1538-3881/aafc33
\bibitem[Gomes \& Papaderos, 2017]{GomPap17} Gomes J.~M., Papaderos P., 2017, A\&A, 603, A63. doi:10.1051/0004-6361/201628986
\bibitem[Gonz{\'a}lez Delgado et al.(2021)]{Gonz21} Gonz{\'a}lez Delgado R.~M., D{\'\i}az-Garc{\'\i}a L.~A., de Amorim A., Bruzual G., Cid Fernandes R., P{\'e}rez E., Bonoli S., et al., 2021, A\&A, 649, A79. doi:10.1051/0004-6361/202039849
\bibitem[Gonz{\'a}lez Delgado et al.(2022)]{Gonz22} Gonz{\'a}lez Delgado R.~M., Rodr{\'\i}guez-Mart{\'\i}n J.~E., D{\'\i}az-Garc{\'\i}a L.~A., de Amorim A., Garc{\'\i}a-Benito R., Mart{\'\i}nez-Solaeche G., Lopes P.~A.~A., et al., 2022, A\&A, 666, A84. doi:10.1051/0004-6361/202244030
\bibitem[Griffin et al., 2010]{Gri10} Griffin M.~J., Abergel A., Abreu A., Ade P.~A.~R., Andr{\'e} P., Augueres J.-L., Babbedge T., et al., 2010, A\&A, 518, L3. doi:10.1051/0004-6361/201014519
\bibitem[Griffith et al., 2011]{Gri11} Griffith R.~L., Tsai C.-W., Stern D., Blain A., Eisenhardt P.~R.~M., Harrison F., Jarrett T.~H., et al., 2011, ApJL, 736, L22. doi:10.1088/2041-8205/736/1/L22
\bibitem[Gupta et al., 2021]{Gupta21} Gupta A., Tran K.-V., Pillepich A., Yuan T., Harshan A., Rodriguez-Gomez V., Genel S., 2021, ApJ, 907, 95. doi:10.3847/1538-4357/abca98
\bibitem[Gupta et al., 2023]{Gupta23} Gupta A., Tran K.-V., Mendel T., Harshan A., Forrest B., Davies R.~L., Wisnioski E., et al., 2023, MNRAS, 519, 980. doi:10.1093/mnras/stac3548
\bibitem[Hermes Team et al., 2017]{Hermes} Hermes Team, Oliver S.~J., Bock J., Altieri B., Amblard A., Arumugam V., Aussel H., et al., 2017, yCat, VIII/103
\bibitem[Iglesias-P{\'a}ramo et al.(2022)]{IP22} Iglesias-P{\'a}ramo J., Arroyo A., Kehrig C., V{\'\i}lchez J.~M., Duarte Puertas S., P{\'e}rez-Montero E., Breda I., et al., 2022, A\&A, 665, A95. doi:10.1051/0004-6361/202243931
\bibitem[Izotov, Thuan, \& Guseva (2012)]{Izo12} Izotov Y.~I., Thuan T.~X., Guseva N.~G., 2012, A\&A, 546, A122. doi:10.1051/0004-6361/201219733
\bibitem[Johnson et al., 2021]{prospector} Johnson B.~D., Leja J., Conroy C., Speagle J.~S., 2021, ApJS, 254, 22. doi:10.3847/1538-4365/abef67
\bibitem[Kehrig et al, 2018]{Keh18} Kehrig C., V{\'\i}lchez J.~M., Guerrero M.~A., Iglesias-P{\'a}ramo J., Hunt L.~K., Duarte-Puertas S., Ramos-Larios G., 2018, MNRAS, 480, 1081. doi:10.1093/mnras/sty1920
\bibitem[Kunth \& Sargent (1986)]{KunSar86}Kunth, D. \& Sargent, W. L. W. 1986, ApJ, 300, 496
\bibitem[Kunth \& {\"O}stlin (2000)]{KunOst00}Kunth, D., \& {\"O}stlin, G. 2000, A\&ARv, 10, 1
\bibitem[Leja et al., 2017]{Lej17} Leja J., Johnson B.~D., Conroy C., van Dokkum P.~G., Byler N., 2017, ApJ, 837, 170. doi:10.3847/1538-4357/aa5ffe
\bibitem[Leja et al., 2017]{Leja17} Leja, J., Johnson, B. D., Conroy, C., van Dokkum, P. G., \& Byler, N. 2017, ApJ, 837, 170
\bibitem[Leja et al., 2018]{Leja18} Leja, J., Johnson, B. D., Conroy, C., \& van Dokkum, P. 2018, ApJ, 854, 62
\bibitem[Leja et al., 2019a]{Leja19a} Leja J., Carnall A.~C., Johnson B.~D., Conroy C., Speagle J.~S., 2019, ApJ, 876, 3. doi:10.3847/1538-4357/ab133c
\bibitem[Leja et al., 2019c]{Leja19b} Leja J., Johnson B.~D., Conroy C., van Dokkum P., Speagle J.~S., Brammer G., Momcheva I., et al., 2019, ApJ, 877, 140. doi:10.3847/1538-4357/ab1d5a
\bibitem[Leja et al., 2020]{Leja20} Leja J., Speagle J.~S., Johnson B.~D., Conroy C., van Dokkum P., Franx M., 2020, ApJ, 893, 111. doi:10.3847/1538-4357/ab7e27
\bibitem[Loeb \& Barkana, 2001]{LoBa01} Loeb A., Barkana R., 2001, ARA\&A, 39, 19. doi:10.1146/annurev.astro.39.1.19
\bibitem[Loose \& Thuan, 1986]{LooThu86} Loose, H. H., \& Thuan, T. X. 1986, Star Forming Dwarf Galaxies and Related Objects (\'Editions Fronti\`eres), 73
\bibitem[Lower et al.(2020)]{Low20} Lower S., Narayanan D., Leja J., Johnson B.~D., Conroy C., Dav{\'e} R., 2020, ApJ, 904, 33. doi:10.3847/1538-4357/abbfa7
\bibitem[Lumbreras-Calle et al.(2021)]{Lum21} Lumbreras-Calle A., L{\'o}pez-Sanjuan C., Sobral D., Fern{\'a}ndez-Ontiveros J.~A., V{\'\i}lchez J.~M., Hern{\'a}n-Caballero A., Akhlaghi M., et al., 2021, arXiv, arXiv:2112.06938
\bibitem[Mainzer et al., 2011]{Main11} Mainzer A., Bauer J., Grav T., Masiero J., Cutri R.~M., Dailey J., Eisenhardt P., et al., 2011, ApJ, 731, 53. doi:10.1088/0004-637X/731/1/53
\bibitem[Martin et al., 2005]{galex} Martin D.~C., Fanson J., Schiminovich D., Morrissey P., Friedman P.~G., Barlow T.~A., Conrow T., et al., 2005, ApJL, 619, L1. doi:10.1086/426387
\bibitem[Mart{\'\i}nez-Solaeche et al.(2021)]{Gines21} Mart{\'\i}nez-Solaeche G., Gonz{\'a}lez Delgado R.~M., Garc{\'\i}a-Benito R., de Amorim A., P{\'e}rez E., Rodr{\'\i}guez-Mart{\'\i}n J.~E., D{\'\i}az-Garc{\'\i}a L.~A., et al., 2021, A\&A, 647, A158. doi:10.1051/0004-6361/202039146
\bibitem[Mart{\'\i}nez-Solaeche et al.(2022)]{Gines22} Mart{\'\i}nez-Solaeche G., Gonz{\'a}lez Delgado R.~M., Garc{\'\i}a-Benito R., D{\'\i}az-Garc{\'\i}a L.~A., Rodr{\'\i}guez-Mart{\'\i}n J.~E., P{\'e}rez E., de Amorim A., et al., 2022, A\&A, 661, A99. doi:10.1051/0004-6361/202142812
\bibitem[Maseda et al., 2013]{Maseda13} Maseda M.~V., van der Wel A., da Cunha E., Rix H.-W., Pacifici C., Momcheva I., Brammer G.~B., et al., 2013, ApJL, 778, L22. doi:10.1088/2041-8205/778/1/L22
\bibitem[Maseda et al., 2014]{Maseda14} Maseda M.~V., van der Wel A., Rix H.-W., da Cunha E., Pacifici C., Momcheva I., Brammer G.~B., et al., 2014, ApJ, 791, 17. doi:10.1088/0004-637X/791/1/17
\bibitem[Oliver et al., 2012]{Oli12} Oliver S.~J., Bock J., Altieri B., Amblard A., Arumugam V., Aussel H., Babbedge T., et al., 2012, MNRAS, 424, 1614. doi:10.1111/j.1365-2966.2012.20912.x
\bibitem[Papaderos et al.(2002)]{P02} Papaderos, P., Izotov, Y.I., Thuan, T.X. et al. 2002 A\&A, 393, 461
\bibitem[Papaderos et al., 1996]{Pap96} Papaderos P., Loose H.-H., Thuan T.~X., Fricke K.~J., 1996, A\&AS, 120, 207
\bibitem[Papaderos et al., 2008]{Pap08} Papaderos P., Guseva N.~G., Izotov Y.~I., Fricke K.~J., 2008, A\&A, 491, 113. doi:10.1051/0004-6361:200810028
\bibitem[Poglitsch et al., 2010]{Pog10} Poglitsch A., Waelkens C., Geis N., Feuchtgruber H., Vandenbussche B., Rodriguez L., Krause O., et al., 2010, A\&A, 518, L2. doi:10.1051/0004-6361/201014535
\bibitem[P{\'e}rez-Montero et al., 2021]{EPM21} P{\'e}rez-Montero E., Amor{\'i}n R., S{\'a}nchez Almeida J., V{\'i}lchez J.~M., Garc{\'i}a-Benito R., Kehrig C., 2021, MNRAS, 504, 1237. doi:10.1093/mnras/stab862
\bibitem[Queiroz et al.(2022)]{Que22} Queiroz C., Abramo L.~R., Rodrigues N.~V.~N., P{\'e}rez-R{\`a}fols I., Mart{\'\i}nez-Solaeche G., Hern{\'a}n-Caballero A., Hern{\'a}ndez-Monteagudo C., et al., 2022, MNRAS.tmp. doi:10.1093/mnras/stac2962
\bibitem[Rahna et al.(2022)]{Rah22} Rahna P.~T., Zheng Z.-Y., Chies-Santos A.~L., Cai Z., Spinoso D., Marquez I., Overzier R., et al., 2022, arXiv, arXiv:2207.00196
\bibitem[Reverte et al., 2007]{Rev07} Reverte D., V{\'\i}lchez J.~M., Hern{\'a}ndez-Fern{\'a}ndez J.~D., Iglesias-P{\'a}ramo J., 2007, AJ, 133, 705. doi:10.1086/510296
\bibitem[Rieke et al., 2004]{MIPS} Rieke G.~H., Young E.~T., Engelbracht C.~W., Kelly D.~M., Low F.~J., Haller E.~E., Beeman J.~W., et al., 2004, ApJS, 154, 25. doi:10.1086/422717
\bibitem[Rossum \& Drake (2009)]{python} Van Rossum, G., \& Drake, F. L. (2009). Python 3 Reference Manual. Scotts Valley, CA: CreateSpace.
\bibitem[Salpeter, 1955]{Sal55} Salpeter E.~E., 1955, ApJ, 121, 161. doi:10.1086/145971
\bibitem[Salvaterra et al., 2011]{Sal11}Salvaterra, R., Ferrara, A., \& Dayal, P. 2011, MNRAS, 414, 847
\bibitem[Schaerer, Contini, \& Pindao (1999)]{Scha99} Schaerer D., Contini T., Pindao M., 1999, A\&AS, 136, 35. doi:10.1051/aas:19991977
\bibitem[Skrutskie et al., 2006]{2mass} Skrutskie M.~F., Cutri R.~M., Stiening R., Weinberg M.~D., Schneider S., Carpenter J.~M., Beichman C., et al., 2006, AJ, 131, 1163. doi:10.1086/498708
\bibitem[Spitzer Science Center (SSC) \& Infrared Science Archive (IRSA),2021]{Spitz} Spitzer Science Center (SSC), Infrared Science Archive (IRSA), 2021, yCat, II/368
\bibitem[Soille, 1999]{MIA} Soille P., Morphological Image Analysis; Principles and Applications, 1999, ISBN 3-540-65671-5
\bibitem[Tang et al., 2019]{Tang19} Tang M., Stark D.~P., Chevallard J., Charlot S., 2019, MNRAS, 489, 2572. doi:10.1093/mnras/stz2236
\bibitem[Thuan \& Martin, 1981]{ThuMar81} Thuan T.~X., Martin G.~E., 1981, ApJ, 247, 823. doi:10.1086/159094
\bibitem[Tran et al., 2020]{Tran20} Tran K.-V.~H., Forrest B., Alcorn L.~Y., Yuan T., Nanayakkara T., Cohn J., Cowley M., et al., 2020, ApJ, 898, 45. doi:10.3847/1538-4357/ab8cba
\bibitem[van der Wel et al., 2011]{Wel11}van der Wel, A., Straughn, A. N., Rix, H. W., et al. 2011, ApJ, 742, 111
\bibitem[Werner et al., 2004]{Wer04} Werner M.~W., Roellig T.~L., Low F.~J., Rieke G.~H., Rieke M., Hoffmann W.~F., Young E., et al., 2004, ApJS, 154, 1. doi:10.1086/422992
\bibitem[Wright et al., 2010]{wise} Wright E.~L., Eisenhardt P.~R.~M., Mainzer A.~K., Ressler M.~E., Cutri R.~M., Jarrett T., Kirkpatrick J.~D., et al., 2010, AJ, 140, 1868. doi:10.1088/0004-6256/140/6/1868
\bibitem[Yang et al., 2016]{Yan16}Yang, H., Malhotra, S., Gronke, M., et al. 2016, ApJ, 820, 130
\bibitem[Withers et al., 2023]{jwst} Withers S., Muzzin A., Ravindranath S., Sarrouh G.~T., Abraham R., Asada Y., Bradac M., et al., 2023, arXiv, arXiv:2304.11181. doi:10.48550/arXiv.2304.11181
\end{thebibliography}
\end{document}